\documentclass[11pt]{article}
\usepackage{epsfig}
\usepackage[usenames]{color}
\renewcommand{\theequation}{\mbox{\arabic{section}.\arabic{equation}}}

\newtheorem{propos}{}[section]
\newcommand{\bprop}{\begin{propos}}
\newcommand{\eprop}{\end{propos}}

\newcounter{Roman}
\addtocounter{Roman}{1}
\newcommand{\prop}[1]{\addtocounter{Roman}{1}
\noindent{\bf \Roman{section}.\arabic{Roman} \ }{\it #1}}
\newcommand{\beq}{\begin{equation}}
\newcommand{\eeq}{\end{equation}}
\newcommand{\bea}{\begin{eqnarray}}
\newcommand{\eea}{\end{eqnarray}}
\newcounter{saveeqn}

\newcommand{\D}{\displaystyle}

\newcommand{\ssc}{\scriptscriptstyle}

\newcommand{\vev}[1]{\Big\langle #1 \Big\rangle}
\newcommand{\dZ}[1]{Z_{\Lambda^{(#1)}}}
\newcommand{\bL}{\Lambda_0}
\newcommand{\dPhi}[1]{\Phi_{\Lambda^{(#1)}}}

\newcommand{\V}{{\cal V}}

\input epsf
\begin{document}

\hfill \vbox{\hbox{}} 
\begin{center}{\Large\bf Confinement for all values of the coupling in 
four-dimensional SU(2) gauge theory}\\[2cm] 
{\bf E. T. Tomboulis\footnote{\sf e-mail: tombouli@physics.ucla.edu}
}\\
{\em Department of Physics, UCLA, Los Angeles, 
CA 90095-1547}
\end{center}
\vspace{1cm}

\begin{center}{\Large\bf Abstract}
\end{center}  
A derivation is given from first principles of the fact that 
the SU(2) gauge theory is in a confining phase for all values 
of the coupling $0 < g < \infty$ defined at lattice spacing 
(UV regulator) $a$, and space-time dimension $d \leq 4$. 
The strategy is to employ approximate 
RG decimation transformations of the potential moving type  
which give both upper and lower bounds on the partition function 
at each successive decimation step. 
By interpolation between these bounds an exact representation of the 
partition function is obtained on progressively coarser lattices.  
In the same manner, one obtains a representation of 
the partition function in the presence of external 
center flux. 
Under successive decimations the flow of the effective action 
in these representations is constrained by that in the upper and 
lower bounds which are easily explicitly computable. 
Confining behavior for the vortex free energy order parameter 
(ratio of partition functions with and without external flux), hence 
`area law' for the Wilson loop, is the result for any initial coupling.  
Keeping the string tension fixed determines the dependence $g(a)$, which 
is such that $g(a)\to 0$ for $a\to 0$.

\vfill 

\pagebreak 

\section{Introduction}

Four-dimensional $SU(N)$ gauge theory at zero temperature is known to be 
in a confining phase for all values of the bare coupling. 
A very large amount of work has been performed over the last decade 
in an effort to isolate the types of configurations in the 
functional measure responsible 
for maintaining one confining phase for arbitrarily weak coupling 
\cite{Rev}, \cite{LAT}. 
Nevertheless, a direct derivation 
of this unique 
feature of $SU(N)$ theories (shared only by 
non-abelian ferromagnetic spin systems in $2$ dimensions)
has remained elusive.

The origin of the difficulty is clear. It is the multi-scale nature of 
the problem: passage from a short distance ordered regime, where weak 
coupling perturbation theory is applicable,  
to a long distance strongly coupled disordered regime, where 
confinement and other collective phenomena emerge. 
Systems involving such dramatic change in physical 
behavior over different scales are hard to treat. Hydrodynamic turbulence, 
involving passage from laminar to turbulent flow, is another 
well-known example, 
which, in fact, shares some striking qualitative features with the 
confining QCD vacuum. 

The natural framework for addressing the problem from first principles 
is a Wilsonian renormalization group (RG) block-spinning procedure bridging 
short to long scales. The use of lattice regularization, i.e. the framework 
of lattice gauge theory (LGT) \cite{W}, is virtually mandatory in this  
context. There is no other known usable 
non-perturbative formulation 
of gauge theory that gives the path integral in closed form 
preserving non-perturbative gauge invariance and positivity of the 
transfer matrix (reflection positivity).  
Attempts at exact blocking constructions towards the 
`perfect action' along the Wilsonian renormalized trajectory \cite{H}, 
however, turn out, not surprisingly, to be exceedingly complicated.

There are, nonetheless, approximate RG decimation procedures that 
can provide bounds on judicially chosen quantities. 
The basic idea in this paper is to 
obtain both upper and lower bounds for the 
partition function and the partition function in the presence of 
external center flux. The bounds are obtained by employing 
approximate decimations of the `potential moving' type \cite{M}, 
\cite{K}, which 
can be explicitly computed to any accuracy by simple algebraic operations.  
This leads to a rather simple construction constraining the behavior of the 
exact partition functions in the presence and in 
the absence of center flux; and, through them, the exact vortex free energy  
order parameter. The latter is the ratio of these two partition functions. 
It is thus shown to exhibit 
confining behavior for all values    
$0 < \beta < \infty$,  of the inverse coupling $\beta=4/g^2$ 
defined at lattice spacing $a$ (UV cutoff). An earlier outline of 
the argument was given in \cite{T1}.

As it will become clear in the following, there are two main ingredients 
here that allow this type of result to 
be obtained. The first is the use of approximate decimations that are 
easily explicitly computable at every step, while correctly reflecting the 
nature of RG flow in the exact theory. The second is to consider 
only partition functions, or (differences of) free energies, rather than 
the RG evolution of a 
full effective action that would allow computation of any observable at 
different scales. This more narrowly focused approach results into 
tremendous simplification compared to a general RG blocking     
construction. 

The presentation is for the most part quite explicit. Some simple 
propositions, mostly containing basic bounds, serve as  
building blocks of the argument. They are enumerated by roman numerals in 
the text below. Most proofs have been relegated to a 
series of appendices so as not to clutter what is essentially a 
simple construction. Only the case of gauge group $SU(2)$ is considered 
explicitly here. The same development, however, can be applied to other 
groups, and, most particularly, to $SU(3)$ which exhibits identical 
behavior under the approximate decimations.

It will be helpful at this point to provide an outline of the 
steps in the argument developed in the rest of the paper. 
In section \ref{DEC}, starting with the pure $SU(2)$  LGT with 
partition function defined on a lattice of spacing $a$, we 
define a class of approximate decimation transformations to a 
coarser lattice of spacing $ba$. 
In section \ref{Z} the resulting partition function on this decimated lattice 
is shown to be an upper bound on the partition function on 
the original lattice. A similar rule can be devised for obtaining a 
partition function on the decimated lattice which gives a lower bound 
on the original partition function. 
One then interpolates between these bounds. 
For some appropriate value of the interpolating parameter, one thus  
obtains an exact 
integral representation of the original partition function. This 
representation is in terms of 
an effective action defined on the decimated lattice of spacing $ab$  
plus a bulk free energy contribution resulting from the blocking $a \to ab$. 
Now, any such interpolation is not unique, 
and it is indeed  expedient to consider different interpolation 
parametrizations.  
The resulting partition function representation is then invariant under 
such parametrization variations in its effective action. 
The other important ingredient is that the effective action in this 
representation is constrained between the effective actions 
corresponding to the upper and lower bound partition functions. 
Iterating 
this procedure in successive decimations, a representation of the 
partition function is  obtained on progressively coarser lattices of 
spacing $a \to ab \to ab^2 \to \cdots \to ab^n$.

In section \ref{TZ} we consider the partition function in the 
presence of external 
center flux. This is the flux of a center vortex, introduced by a 
$Z(2)$ `twist' in the action, and rendered topologically stable by winding 
around the lattice torus. The decimation-interpolation  procedure 
just outlined for the partition function can be applied also in the 
presence of the external flux. A  representation of the twisted 
partition function on progressively coarser lattices can then be 
obtained in a completely analogous manner.

The ratio of the twisted to the untwisted partition function is the 
vortex free energy order parameter. Its behavior as a function 
of the size of the system characterizes the system's possible phases. 
By known correlation inequalities it can, furthermore, be related to the 
Wilson and t'Hooft order parameters. Our representations 
of the twisted and untwisted partition functions may now be used 
to represent the ratio (section \ref{Z-/Z}).    
One may exploit the parametrization invariance of these representations 
to ensure that the bulk free energy contributions resulting in each 
decimation step $ab^{m-1} \to ab^m$  explicitly cancel between numerator and 
denominator in the ratio. One is then left with a representation 
of the vortex free energy solely in terms of an effective action 
defined on a lattice of spacing $ab^n$.

Now this effective action is constrained by 
the effective actions corresponding to the upper and lower bounds. 
The latter are easily explicitly computable by straightforward iteration 
of the potential-moving decimation rules. Under successive 
transformations they flow, for space-time dimension $d\leq 4$ and 
any original coupling $g$ defined at 
spacing $a$, to the strong coupling regime.  This is the regime where the 
coefficients in the character expansion of the exponential of the action 
become sufficiently small for the strong coupling cluster expansion to 
converge. Confining behavior is the immediate result for the 
vortex free energy, and, hence, `area law' behavior for the Wilson loop 
(section \ref{CONf}). 

As it is well-known the theory contains only one free parameter, a 
physical scale which is conventionally taken to be (some multiple of) 
the string tension. This fact comes out in a natural way in the 
context of RG decimations, as we will see in the following. 
Fixing this scale then determines the dependence $g(a)$. 
The fact that $g(a)\to 0$ as $a\to 0$ is an essentially 
qualitative consequence of the flow exhibited by the decimations. 

Some concluding remarks are given in section \ref{SUM}.

\section{Decimations} \label{DEC}
\setcounter{equation}{0}
\setcounter{Roman}{0}

We work on a hypercubic lattice $\Lambda \subset {\rm\bf Z}^d$ of 
length $L_\mu$ in the $x^\mu$-direction, $\mu=1,\ldots ,d$, 
in units of the lattice spacing $a$. Individual bonds, plaquettes, 
3-cubes, etc are generically denoted by $b$, $p$, $c$, etc. More 
specific notations such as $b_\mu$ or $p_{\mu\nu}$  are 
used to indicate elementary $m$-cells of particular orientation. 
We use the standard framework and common notations of LGT with 
gauge group $G$. Group elements are generically denoted by $U$, 
and the bond variables by $U_b \in G$. In this paper we take 
$G=SU(2)$. 
 
We start with some appropriate plaquette action $A_p$ defined 
on $\Lambda$, which, for definiteness, is taken to be 
the Wilson action 
\beq
A_p(U_p,\beta) ={\beta\over 2}\;{\rm Re}\,\chi_{1/2}(U_p) \;, \qquad 
U_p=\prod_{b\in \partial p} U_b \;,\label{Wilson}
\eeq
with $\beta=4/g^2$ defining the lattice coupling $g$.   
The character expansion 
of the exponential of the plaquette action function is given by   
\beq 
\exp \left(A_p(U,\beta)\right) 
   = \sum_j\;d_j\, F_j(\beta)\,\chi_j(U) \label{exp} 
\eeq
with Fourier coefficients: 
\beq
 F_j(\beta) = \int\,dU\;
\exp \left(A_p(U,\beta)\right) \,{1\over d_j}\,\chi_j(U)\;.\label{Fourier}
\eeq
Here $dU$ denotes Haar measure on $G$, and $\chi_j$ the 
character of the $j$-th representation of dimension $d_j$. 
So, for SU(2), the only case considered explicitly here, all 
characters are real, $j=0, {1\over 2}, 1, 
{3\over 2}, \ldots$, and $d_j=(2j+1)$. (\ref{Fourier}) implies that 
$F_0\geq F_j$, all $j\not=0$. Explicitly, one finds 
\beq 
F_j(\beta) = {2\over \beta}\, I_{d_j}(\beta) \, 
\eeq
in terms of the modified Bessel function $I_\nu$.  
 
It will be convenient to work in terms of normalized coefficients: 
\beq 
c_j(\beta) = {\D F_j(\beta) \over \D F_0(\beta)} \;, \label{ncoeffs}
\eeq
so that   
\bea
\exp \left(A_p(U,\beta)\right)  &=&  F_0\,\Big[\, 1 + \sum_{j\not= 0} 
d_j\,c_j(\beta)\,\chi_j(U)\,
 \Big] \nonumber \\
    & \equiv & F_0\;f_p(U,\beta)\,.  \label{nexp} 
\eea 
The (normalized) partition function on lattice $\Lambda$ is then  
\beq
Z_\Lambda(\beta) = 
\int dU_\Lambda\;\prod_{p\in \Lambda}\,f_p(U_p,\beta)\equiv 
\int\,d\mu_\Lambda^0
\;,\label{PF1}
\eeq 
where $dU_\Lambda\equiv \prod_{b\in \Lambda}dU_b$, and expectations 
are computed with the measure $d\mu_\Lambda = 
d\mu_\Lambda^0 / Z_\Lambda(\beta)$. 
 
The action (\ref{Wilson}) is  such that 
\beq
F_j (\beta)\geq 0\;, \qquad \mbox{hence}\quad 1\geq c_j(\beta)
\geq 0\qquad\quad \mbox{all}\quad j \;,  
\eeq 
which implies that the measure defined by (\ref{PF1}) satisfies reflection 
positivity (RP) both in planes without sites and in planes with sites. 
Note that $\lim_{\beta\to \infty}c_j(\beta)=1$.  

Let $\Lambda^{(n)}$ be the hypercubic lattice of spacing $b^na$, 
with integer $b\geq 2$, and  
$Z_{\Lambda^{(n)}}(\{c_j(n)\})$ denote a  partition 
function of the form (\ref{PF1}) defined on 
$\Lambda^{(n)}$ in terms of 
some given set of coefficients $\{c_j(n)\}$: 
\bea
Z_{\Lambda^{(n)}}(\{c_j(n)\}) & = & 
  \int dU_{\Lambda^{(n)}} \prod_{p\in \Lambda^{(n)}}
\Big[\, 1 + \sum_{j\not= 0} d_j \, 
c_j(n)\chi_j(U_p)\,\Big] \nonumber \\
& \equiv & \int dU_{\Lambda^{(n)}} \prod_{p\in \Lambda^{(n)}}\,f_p(U_p,n)
\equiv \int\,d\mu_{\Lambda^{(n)}}^0
\,,\label{PF2}
\eea 
where $dU_{\Lambda^{(n)}}\equiv \prod_{b\in \Lambda^{(n)}}dU_b$. 
We also employ the notations 
\beq
g_p(U,n) \equiv f_p(U,n) -1 = \sum_{j\not= 0} d_j\, c_j(n)
\, \chi_j(U) \;,\label{g}
\eeq
and $\| \cdot\|$ for the $\|\cdot\|_\infty$-norm:  
\beq
\|g(n)\| = \sum_{j\not=0} d_j^2\,  c_j(n) \,.\label{gnorm}
\eeq

One has the simple but basic result: 

\prop{For $Z_{\Lambda^{(n)}}(\{c_j(n)\})$ given by (\ref{PF2}) 
with $c_j(n) \geq 0$ for all $j$, and periodic boundary conditions, 

(i) $\dZ{n}(\{c_j(n)\})$ is an increasing function of each 
$c_j(n)$: 
\beq 
\partial \dZ{n}(\{c_i(n)\}) / \partial c_j(n) \geq 0 \; ;\label{PFder0}
\eeq

(ii) 
\beq 
Z_{\Lambda^{(n)}}(\{c_j(n)\}) \geq \Big[\,1 + \sum_{j\not=0} d_j^2\,
c_j(n)^6 \,\Big]^{|\Lambda^{(n)}|} \; .\label{PFlowerb1}
\eeq  } 
(\ref{PFder0}) is an immediate consequence of RP in planes without sites. 
The proof of (\ref{PFlowerb1}), also based on RP, is given in Appendix A. 
Strict inequality in fact holds in (\ref{PFder0}) and (\ref{PFlowerb1}),   
with equality only in the trivial case where all $c_j(n)$'s vanish. 
In particular, one has 
\beq  
Z_{\Lambda^{(n)}}(\{c_j(n)\}) \; > \; 1 \,.\label{PFlowerb2}
\eeq
Simple as (\ref{PFlowerb2}) is, it is not trivial, as it requires 
non-negativity of $c_j(n)$'s, and will be useful in the following.

\subsection{Construction of decimation transformations} \label{DEC1}

To perform an RG transformation $a \to ba$,  
the lattice is partitioned into $d$-dimensional 
decimation cells of side length $ba$. Various approximate 
decimation transformations may be devised involving the  
`weakening', i.e. decreasing the $c_j$'s  of interior plaquettes, while 
compensating by `strengthening', i.e. increasing $c_j$'s  of 
boundary plaquettes of each cell. The simplest such scheme \cite{M}, 
which is  adopted in the following,  implements  
complete removal of interior plaquettes. This may be pictured \cite{K}
as moving the potentials due to interior plaquette interactions 
to the boundary.  

\begin{figure}[ht]
\resizebox{15cm}{!}{\input{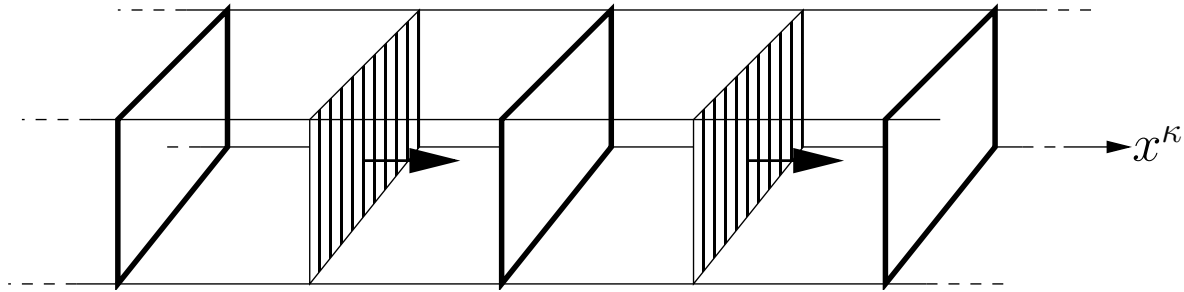}}  
\caption{Basic plaquette moving operation, $b=2$ \label{Dec1fig}}
\end{figure}

This `potential moving' may be performed  as the composition of 
elementary steps. 
The elementary potential moving  step is defined in terms of 
a $3$-dimensional cell of side length $ba$ in a given 
decimation direction, say the $x^\kappa$-direction, 
and length $a$ in the other two directions $\mu$, $\nu$. 
Two such $3$-cells adjacent along the 
$\kappa$-direction are shown in Figure~\ref{Dec1fig}. The $(b -1)$ 
interior plaquettes in each cell perpendicular 
to $x^\kappa$ (shaded) are removed, i.e. 
\beq 
A_p(U_p) \to  0 \label{potmove1}
\eeq
for the action at their original location, and displaced (arrows) in the
positive $x^\kappa$ direction to the position of the corresponding 
plaquette (bold) on the cell boundary. There the displaced interior 
plaquettes are combined with the boundary plaquette 
into one plaquette $p$ with action   
`renormalized' by some appropriate  amount\footnote{One 
may take this renormalization factor to depend on the move direction, 
but we need not consider these more general transformations here.}
$\zeta_0$:
\beq 
A_p(U) \to  \zeta_0\,A_p(U) \;.\label{potmove2}
\eeq 
A complete transformation consists of performing this elementary operation 
successively in every lattice direction $\kappa=1, \ldots, d$ 
in such a way that eventually one is left only with plaquette interactions 
on a lattice of spacing $ba$. 
In practice, there is no reason for a choice 
other than $b=2$, but, for clarity, we keep general (integer) $b$.  
The result of a complete transformation 
is given by equations (\ref{RG1})-(\ref{RG5}) below, to which a reader may 
turn directly.

To describe this process in more detail, let  the lattice be 
partitioned into $d$-dimensional hypercubic decimation cells $\sigma^d$ 
of side length $b a$ in each lattice direction. Plaquettes interior 
to a $\sigma^d$ are defined as those  not wholly contained in its 
$(d-1)$-dimensional boundary $\partial \sigma^d$. 
Consider the effect 
of successive application of the elementary moving operation to plaquettes 
of fixed orientation, say $[\mu\nu]$. 
There are $(d-2)$ normal directions $\kappa_i \not= \mu,\nu$, 
$i=1, \cdots, d-2$, 
in which a plaquette $p_{\mu\nu}$ can be moved. 
Interior $p_{\mu\nu}$'s in each $\sigma^d$ are first moved 
to the cell boundary $\partial \sigma$   
in groups of $(b-1)$ parallel plaquettes,  
along , say, the positive $\kappa_1$-direction (as in Figure 1). They end up 
in the face $\sigma^{(d-1)}_{\kappa_1} \subset \partial \sigma^d$  
perpendicular to the $\kappa_1$-axis. 
There each group is identified with the plaquette present at that 
location and merged in one plaquette $p_{\mu\nu} \in 
\sigma^{(d-1)}_{\kappa_1}$ 
with a `renormalized' action (\ref{potmove2})).
Similarly, $p_{\mu\nu}$ plaquettes in each face 
$\sigma^{(d-1)}_{\kappa_i}\subset \partial \sigma^d$, 
with $i\not= 1, \mu,\nu$ are moved along the $\kappa_1$-axis in 
groups of $(b-1)$  to the face $\sigma^{(d-2)}_{\kappa_1\kappa_i}
\subset \partial \sigma^{(d-1)}_{\kappa_i}$  
normal to the $\kappa_1$ and $\kappa_i$ directions, where they are  
merged and renormalized. 

There are now $(d-3)$ directions inside the $(d-1)$-dimensional face 
$\sigma^{(d-1)}_{\kappa_1}$ in  
which a $[\mu\nu]$-plaquette can move. 
Thus in proceeding to apply  the elementary moving operation successively 
in all directions, the once-moved-renormalized $p_{\mu\nu}$'s 
in $\sigma^{(d-1)}_{\kappa_1}$ are next 
moved, in groups of $(b-1)$ plaquettes in the positive $\kappa_2$-direction, 
to the face $\sigma^{(d-2)}_{\kappa_1\kappa_2} \subset \partial 
\sigma^{(d-1)}_{\kappa_1}$. Similarly,  
the once-moved-renormalized $p_{\mu\nu}$'s inside a face 
$\sigma^{(d-2)}_{\kappa_1\kappa_i}$ are moved provided $\kappa_2$ is among the 
$(d-4)$ available directions normal to a $[\mu\nu]$-plaquette inside  
$\sigma^{(d-2)}_{\kappa_1\kappa_i}$. 

Continuing this process in the remaining directions 
$\kappa_i$, $i=3,\ldots,(d-2)$, the set of 
$[\mu\nu]$-plaquettes on the initial lattice ends up in the 
$2$-dimensional faces 
$\sigma^2_{\kappa_1\kappa_2\ldots\kappa_{(d-2)}} \subset \partial 
\sigma^3_{\kappa_1\kappa_2\ldots\kappa_{(d-3)}} 
\linebreak 
\subset \cdots \subset 
\partial \sigma^{(d-1)}_{\kappa_1}$.

The above process, described for plaquettes of one 
fixed orientation $[\mu\nu]$, is carried out for each of the 
$d(d-1)/2$ possible choices of plaquette orientation \cite{K}. 

The end result of the process is then a lattice having   
elementary $2$-faces of side length $b a$, each tiled by 
$b^2$ plaquettes of side length $a$.  
The action of each of these $b^2$ plaquettes has been 
renormalized according to (\ref{potmove2}) by a total factor of 
\beq 
\zeta_0^{(d-2)} \equiv \zeta 
\;. \label{totalren}
\eeq   
This is expressed by (\ref{RG5}) below.

The integrations over the bonds interior to each   
$2$-face of side length $ba$ are now carried out. This merges the 
$b^2$ tiling plaquettes into a single plaquette of side length $ba$.  
These integrations are exact   
and do not change the value of the partition 
function that resulted after the completion of the plaquette moving 
operations. We, however, allow further renormalizing the result of these 
integrations by introducing, in addition to $\zeta_0$, another parameter,  
$r$ (cf. (\ref{RG2}) below).   
This completes the decimation transformation to a hypercubic 
lattice of spacing spacing $ba$.

The important feature of this decimation transformation is 
that it preserves the original one-plaquette form of the action, 
so the result can again be represented in the form (\ref{nexp}).  
The transformation rule for successive decimations 
\bea 
& & a\, \to b\, a \,\to\, b^2 a \to\, \cdots \to\,b^{n-1} a
 \to \,b^n a \to \cdots\nonumber \\
 & & \Lambda \to \Lambda^{(1)} \to \Lambda^{(2)} \to \cdots \to 
\Lambda^{(n-1)} \to \Lambda^{(n)} \to \cdots\;,\nonumber  
\eea
is then: 
\beq
f_p(U,n-1)\to F_0(n)\,f_p(U,n) = F_0(n)\,\Big[ 1 + \sum_{j\not= 0} 
d_j\,c_j(n)\,\chi_j(U) \Big]  \,.\label{RG1} 
\eeq   
The $n$-th step coefficients $F_0(n)$, $c_j(n)$ are obtained from 
the coefficients $c_j(n-1)$ of the previous step by  
\beq
c_j(n) = \hat{c}_j(n)^{b^2 r}\; , \label{RG2}
\eeq
\beq 
F_0(n) = \hat{F}_0(n)^{b^2} \label{RG3}
\eeq
where 
\beq
\hat{c}_j(n)\equiv \hat{F}_j(n)/\hat{F}_0(n) \leq 1  \;, \qquad j\not= 0\;,
\label{RG4}
\eeq
and
\beq
\hat{F}_j(n)= \int\,dU\;\Big[\,f(U,n-1)\,\Big]^\zeta\,
{1\over d_j}\,\chi_j(U) 
\; .  \label{RG5}
\eeq
The $n=0$ coefficients are the coefficients 
$c_j(\beta)$ on the original lattice $\Lambda$.   
(\ref{RG5}) encodes the end result of the plaquette moving - renormalization 
operations described above, with $\zeta$ of the form (\ref{totalren});  
and (\ref{RG2}), (\ref{RG3}) that of the 
subsequent 2-dimensional integrations, and further renormalization by 
the parameter $r$. 

It is easily seen that $f_p(U, n) \;>\; 0$ given that this holds for $n=0$ 
(cf. (\ref{exp}), (\ref{nexp})). 
The effective plaquette action 
on lattice $\Lambda^{(n)}$ of spacing $b^n a$ 
is then
\bea 
f_p(U, n) & = & \Big[\, 1 + \sum_{j\not= 0} d_j\,c_j(n)\, \chi_j(U)\,
 \Big]  \label{nexp1} \\
& \equiv & \exp\Big(\, A_p(U, n)\,\Big)   \;, \label{actdef}
\eea 
with effective couplings defined by the character expansion  
\beq 
A_p(U, n) = \beta_0(n) + \sum_{i\not= 0} \beta_i(n)\,d_i\,\chi_i(U) \;. 
\label{effact} 
\eeq

A point on notation. In the above we used the notations $F_0(n)$, 
$c_j(n)$, $\beta_j(n)$, etc, which do not display the full set of 
explicit or implicit dependences of these quantities. Thus, a more complete 
notation is:
\bea
c_j(n) &=& c_j(\,n,b,\zeta,r,\{c_j(n-1)\}\,)   \nonumber \\
F_0(n) &=& F_0(\,n,b,\zeta,\{c_j(n-1)\}\,) \label{short-hand}
\eea
Dependence on the original coupling $\beta$ comes, of course, 
iteratively through the coefficients $\{c_j(n-1)\}$ of the preceding step.     
Because of the iterative nature of many of the arguments in this paper 
several  explicit and implicit dependences propagate to most of the 
quantities used in the following. To prevent notation from getting out of 
hand we generally employ short-hand notations such as those on the l.h.s. 
of (\ref{short-hand}), unless specific reference to particular 
dependences is required.

The resulting partition function after $n$ such 
decimation steps is: 
\beq 
Z_\Lambda(\beta, n) = 
\prod_{m=1}^n  F_0(m)^{|\Lambda|/b^{md}}\; Z_{\Lambda^{(n)}}(\{c_j(n)\})
\; ,\label{PF2a}
\eeq
with  $Z_{\Lambda^{(n)}}(\{c_j(n)\})$ of the form (\ref{PF2}) and 
coefficients (\ref{short-hand}) resulting after $n$ steps according to 
(\ref{RG2}) - (\ref{RG5}). 
The bulk free energy density resulting 
from decimating from scale $a$ to $b^n a$ is then  
$\sum_{m=1}^n \ln F_0(m) /b^{md}\,$, 
each term in this sum representing the contribution from 
$b^{(m-1)}a \to b^m a$ as specified by (\ref{RG1}).  
The partition function (\ref{PF2a}) is, of course, not equal to 
the original partition function $Z_\Lambda(\beta)$ of (\ref{PF1}) 
since the decimation transformation is not exact. How they are related 
will be addressed below.

\subsection{Some properties of the decimation transformations}\label{DEC2}

The transformation rule specified by (\ref{RG1})-(\ref {RG5}) is 
meaningful for real positive $\zeta$. Here, however, a basic 
distinction can be made. 
As it is clear from (\ref{RG5}), 
for {\it integer} $\zeta$ the important property of positivity of the Fourier 
coefficients in (\ref{RG1}) is maintained at each decimation step:   
\beq
F_0(n)\geq 1\;, \qquad 1\geq c_j(n)\geq 0 \qquad \qquad
(\mbox{integer} \ \zeta) \;. \label{+c}
\eeq 
This means that reflection positivity 
is maintained at each decimation step. 
This clearly is not  guaranteed to be 
the case for non-integer $\zeta$. 
Thus non-integer $\zeta$ results in transformations that, in general,  
violate the reflection positivity of the theory (assuming a 
reflection positive original action). 

It is important in this connection that,  
after each decimation 
step, the resulting action retains the original one-plaquette form,  
but will generally contain {\it all} representations in (\ref{effact}). 
Furthermore, among the effective couplings 
$\beta_j(m)$ negative ones will occur. 
These features are present in general, even after a single decimation 
step $a\to ba$ starting, as we did, with the single (fundamental)
representation Wilson action (\ref{Wilson}).  For integer $\zeta$, however, 
the resulting effective action (\ref{effact}), even in the presence of some 
negative couplings, still defines a reflection positive measure, 
since, as just noted, the expansion of its exponential (\ref{nexp1}) gives 
positive coefficients (\ref{+c}).

It is also worth noting that, given a set of initial coefficients 
(\ref{ncoeffs}), the transformation 
rule (\ref{RG2}) - (\ref{RG5}) with integer $\zeta$ can be explicitly 
evaluated, to any desired accuracy, by purely algebraic operations, namely  
repeated application of the KG reduction rule 
\beq 
\chi_i\,\chi_j =\sum_{k=|i-j|}^{i+j} \chi_k \label{KG}
\eeq 
in (\ref{RG5}) and character orthogonality -- no actual integrations 
need be carried out.

The choice (cf. (\ref{totalren}))
\beq 
\zeta_0 = 1 +(b -1 ) \qquad \Longrightarrow \qquad 
\zeta=b^{d-2} 
\label{MKz}
\eeq 
is special. It increases the couplings of  
receiving plaquettes, at each basic moving step, 
by an amount exactly equal to that of the 
corresponding displaced plaquettes. This, together with $r=1$, is 
essentially the original 
choice in \cite{M} as reformulated in \cite{K}, and will be referred to as 
MK decimation. It will be important in the following.\footnote{It is worth 
noting in this context that in numerical investigations 
of the standard MK recursions in gauge theories \cite{NT-BGZ} 
fractional $b$, ($1< b <2$), which by (\ref{MKz}) corresponds to 
non-integer $\zeta$, has often been used.}

There are various other interesting properties of the decimations 
that can be derived from (\ref{RG2}) - (\ref{RG5}).  
The following one is particularly important. The norm  (\ref{gnorm}) 
of the coefficients obtained by application of (\ref{RG2}) - (\ref{RG5}) 
with integer $\zeta\geq 1$ and $r=1$ satisfies (Appendix D): 
\beq
||g(n+1)|| \leq \Big[\,\zeta\,||g(n)||\,\Big]^{b^2}
\Big[\, 1+ ||g(n)||\,\Big]^{(\zeta-1)b^2} \,.\label{gnormrecur}
\eeq

Assume now that 
\beq 
||g(n)|| \leq \exp (- C_n)\,. \qquad \quad C_n > 0 \,, \label{gnormU1}
\eeq
for some $n$. Then 
\beq
||g(n+1)|| \leq \Big[\,\zeta\,||g(n)||\,\Big]^{b^2} \exp \left(\,
(\zeta-1)\, b^2\,\right)
\leq \exp \Big[ -(\, C_n - k\,) b^2 \Big] \,, \label{gnormrecurU1}
\eeq
where $k= \ln \zeta + (\zeta-1)$. The recursion 
\beq 
C_{n+1} = C_n b^2 - k b^2  \label{gnormrecurU2}
\eeq
gives
\beq
C_{n+m} = \Big[\, C_n - {b^2 k\over b^2-1}\,\Big] b^{2m} + {b^2 k\over b^2-1}
\,.\label{gnormsoln}
\eeq 
\prop{If for some $n$ the norm of coefficients (\ref{gnorm}) obeys 
(\ref{gnormU1}) with 
\beq
C_n > {b^2 k\over b^2-1}\, , \label{gnormU2}
\eeq 
then, under iteration of the decimation transformation 
(\ref{RG2}) - (\ref{RG5}), $||g(n+m)||\to 0$ as  
$m\to \infty$ according to (\ref{gnormsoln}). }\\
This fall-off behavior is immediately 
recognizable as ``area-law''.

If one assumes that $c_j(n)$ are small enough so that the theory 
is within the strong coupling regime, this behavior can be immediately 
deduced for the leading coefficient $c_{1/2}(n)$ directly from 
(\ref{RG2}) - (\ref{RG5}):  
\beq
c_{1/2}(n+1) = c_{1/2}(n)^{b^2} 
\exp \Big(\,[\,\ln \zeta + O(c_{1/2}(n))\,]\,b^2\, \Big)\,. \label{RGstrong} 
\eeq
The result (\ref{gnormrecurU1}) gives then 
an estimate of the corrections due to all higher representations. 
What is noteworthy here, however, is that the condition (\ref{gnormU2}) 
is rather weaker than the commonly stated conditions  for being inside the
convergence radius of the strong coupling cluster expansion (cf. 
section \ref{CONf}).

We note two further properties of the decimation 
transformations (\ref{RG1}) - (\ref{RG5}). The first is that with $r=1$ 
they become exact in space-time dimension $d=2$ since then, from 
(\ref{totalren}),  $\zeta=1$. 
The second is that, with $\zeta=b^{(d-2)}$, vanishing coupling 
$g=0$ is a fixed point in any $d$, i.e. MK decimation is 
exact at zero coupling. This follows simply from the fact that 
\[ \lim_{\beta\to \infty} \Big[\int d\nu(x)\;  e^{\beta f(x)} \Big]^{1/\beta} 
=\mbox{ess. sup}\   e^{f(x)} \equiv \| e^f\|  \] 
for any normalized measure $d\nu(x)$. Applying this to the  result of 
performing the plaquette moving operation starting from 
(\ref{nexp}), and with $p^\prime\in \Lambda$ labeling the plaquettes 
tiling the plaquettes $p\in \Lambda^{(1)}$, one has 
\bea
& & \lim_{\beta\to \infty} \Bigg[\int dU_\Lambda \,
\prod_{p\in \Lambda^{(1)}} 
\prod_{p^\prime \in p}\exp\Big(\beta b^{(d-2)}\,{1\over 2}
\chi_{1/2}(U_{p^\prime})\Big) \Bigg]^{1/\beta}  \nonumber \\
& = & 
\prod_{p\in\Lambda^{(1)}} \Big\|\exp \Big(b^{(d-2)}{1\over2} \chi_{1/2}\Big)
\Big\|^{b^2} = e^{|\Lambda|} = 
\lim_{\beta\to \infty} \Bigg[\int dU_\Lambda\,\prod_{p\in \Lambda} 
\exp\Big(\beta \,{1\over 2}
\chi_{1/2}(U_p)\Big) \Bigg]^{1/\beta} \;.\qquad
\eea
This clearly holds also for $r\not= 1$, as is evident from 
the fact that $\lim_{\beta\to \infty} c_j(\beta)=1$. This fixed point is 
easily seen to be unstable.  

\section{Partition function} \label{Z}
\setcounter{equation}{0}
\setcounter{Roman}{0}

Since our decimations are not exact RG transformations, the partition 
function does not in general remain invariant under them. 
The subsequent development hinges on the following two basic propositions 
that relate partition functions under such a decimation.

\subsection{Upper and lower bounds}\label{u-lPF} 
Consider a partition $Z_{\Lambda^{(n-1)}}$ on lattice $\Lambda^{(n-1)}$ of the 
form (\ref{PF2}) given in terms of some set of coefficients $\{c_j(n-1)\}$. 
Apply a decimation transformation (\ref{RG1}) - (\ref{RG5}) performed with 
$\zeta=b^{(d-2)}$. Denote the resulting coefficients by $c_j^U$, $F_0^U$, 
i.e.   
\bea
c_j^U(n,r) & \equiv & c_j(\, n,b,\,\zeta=b^{(d-2)}, r, \{c_j(n-1)\}\,)  
\label{upperc}\\
F_0^U(n) & \equiv & F_0( \, n,b,\, \zeta=b^{(d-2)},  \{c_j(n-1)\}\,) 
\label{upperF} \;. 
\eea 
Note that 
\beq 
c_j^U(n,r) = c_j^U(n,1)^r \, .
\label{upperc1}
\eeq 
 
\prop{ 
For $Z_{\Lambda^{(n-1)}}$ of the form (\ref{PF2}),  
a decimation transformation (\ref{RG1}) - 
(\ref{RG5}) with $\zeta=b^{d-2}$ and $0 < r\leq 1$  results in an upper 
bound on $Z_{\Lambda^{(n-1)}}$: 
\beq
Z_{\Lambda^{(n-1)}}(\{c_j(n-1)\})\,  \leq \,  F_0^U(n)^{|\Lambda^{(n)}|}\, 
Z_{\Lambda^{(n)}}(\{c_j^U(n,r)\})\;.\label{U}
\eeq 

The r.h.s. in (\ref{U}) is a monotonically 
decreasing function of $r$ on $0 < r\leq 1$. 
}

Given partition function $Z_{\Lambda^{(n-1)}}$ on lattice 
$\Lambda^{(n-1)}$ of the form (\ref{PF2}) in terms of some set of 
coefficients $\{c_j(n-1)\}$, let   
\bea 
c_j^L(n) & \equiv & c_j(n-1)^6 \label{lowerc1}\\
F_0^L(n) & \equiv & 1 \;.\label{lowerF1}
\eea 

\prop{ 
For $\dZ{(n-1)}$, $\dZ{n}$ of the form (\ref{PF2}): 
\beq 
Z_{\Lambda^{(n)}}(\{c_j^L(n)\}) \, \leq \, Z_{\Lambda^{(n-1}}(\{ 
c_j(n-1)\}) \;.  \label{L}
\eeq   
}

The proof of III.1 is given in Appendix A,    
where somewhat stronger results than (\ref{U})  
are actually obtained. III.2 is a corollary of 
(\ref{PFlowerb1}) (Appendix A). For the argument in the rest of this paper, 
the precise form of the lower bound is in fact not important.  
By II.1(i) a further lower bound is   obtained by replacing 
$c_j^L(n)$ in III.2 by, for example,   
\beq 
c_j^L(n) \equiv  c_j(n-1)^6 \,c_j^U(n,r) \label{lowerc2}
\eeq
since $0\leq c_j^U(n,r)\leq 1$. Another choice is to 
simply set 
\beq 
c_j^L(n)=0 \,,\label{lowerc3}
\eeq 
which is a restatement of (\ref{PFlowerb2}). 

A related lower bound, which, in analogy to the upper bound in III.1,   
can be formulated directly in terms of the 
transformations (\ref{RG1}) - (\ref{RG5}), is obtained by taking 
$c_j^L(n)$ in III.2 to be given by: 
\bea
c_j^L(n) & \equiv & c_j(\, n,b,\,\zeta=1, r=1, \{c_j(n-1)\}\,) \nonumber\\
           & = & c_j(n-1)^{b^2} \,, 
\label{lowerc4}\\
F_0^L(n) & \equiv & F_0( \, n,b,\, \zeta=1,  \{c_j(n-1)\}\,) \nonumber \\
           & = & 1 \; .\label{lowerF2}
\eea  
With this choice of $c_j^L(n)$, note that III.1 - III.2  imply the 
fact that the decimations (\ref{RG1}) - (\ref{RG4}) 
become exact for $d=2$ and $r=1$. 
III.1 says that, after removal of interior plaquettes,  modifying the 
couplings of the remaining plaquettes  
by taking $\zeta=b^{d-2}$   (and $r\leq 1$) 
results into overcompensation. 
III.2 says that decimating plaquettes while leaving the couplings of the 
remaining plaquettes unaffected ($\zeta=1$, $r=1$) results in 
undercompensation. 
The proof of III.2 for $c_j^L(n)$ given by (\ref{lowerc4})
is similar to that of II.1,  
but need not be given here, since the weaker  
bounds above will suffice.

In the following it will in fact be more convenient to take (\ref{lowerc2}) or 
(\ref{lowerc3}) for the definition of the lower bound coefficients 
$c_j^L(m)$.  Use of the stronger lower bounds 
above may be preferable 
for numerical investigations, but does not contribute anything further  
to the argument in this paper.   

III.1 and III.2 give  upper and lower bounds on the partition 
function after a decimation step. 
It is then natural to interpolate between these bounds.

\subsection{Interpolation between upper and lower bounds}\label{interbounds}

Introducing a parameter $\alpha \in [0,1]$, we define 
coefficients $\tilde{c}_j(m,\alpha,r)$ 
interpolating between $c_j^L$ at $\alpha=0$ and $c_j^U$ 
(\ref{upperc}) at $\alpha=1$: 
\beq
\tilde{c}_j(m,\alpha, r) = (1-w(\alpha))\, c_j^L(m) + 
w(\alpha)\, c_j^U(m,r) \;, \quad \qquad 0 < r \leq 1. 
\label{interc1}
\eeq 
with 
\beq
w(0)=0\;, \qquad \quad w(1)=1\;,  \quad \qquad w^\prime(\alpha) >  0 
\;. \label{interc2} 
\eeq 
For example, 
\beq
w(\alpha) = {e^\alpha-1\over e-1}  \label{w}
\eeq 
There is clearly a variety of other choices than (\ref{interc1}) for these 
interpolating coefficients. 
We always require that
\beq 
\partial\, \tilde{c}_j(m,\alpha,r) /\partial \,\alpha > 0  \;,
\label{interc3}
\eeq 
which is satisfied by (\ref{interc1}) - (\ref{interc2}).

Similarly, we define coefficients  
interpolating between (\ref{lowerF1}) and (\ref{upperF}). For our purposes 
it will be convenient to take 
\beq 
\tilde{F}_0(m,h, \alpha,t) = F_0^U(m)^{h_t(\alpha)} \label{interF1} \;, 
\eeq
where $h_t$ denote a family of monotonically increasing smooth functions 
of $\alpha$, labeled by a parameter 
$t \in [t_a,t_b]$, and such that 
\beq 
h_t(0)=0\;, \qquad h_t(1)=1 \,. \label{hlimits}
\eeq
We write $h_t(\alpha) \equiv h(\alpha,t)$. Examples 
are\footnote{Supplementing 
these definitions at $\alpha=0$ as needed is understood. Thus, $h(\alpha,t)=  
0$ on $\alpha \leq 0$ in the first example in (\ref{h}); and 
standard smoothing in the second example: replace $\alpha$ in $h$ by 
$g_\epsilon(\alpha)=\int \rho_\epsilon(\alpha-x) g(x)dx$,  where 
$g(x)=x$ for $x>0$, $g(x) =0$ for $x\leq 0$, and $\rho_\epsilon(x)$
is $C^\infty$, has support inside $|x|^2\leq \epsilon^2$ and 
satisfies $\rho_\epsilon \geq 0$ and $\int \rho_\epsilon =1$. }    
\bea 
h(\alpha,t) & = & \exp \left( -\, \sigma(t)\,{1-\alpha\over \alpha}\right)\;, 
\qquad h(\alpha,t) =\alpha^{\sigma(t)}\, , \qquad 
h(\alpha,t) = \tanh\,({\ \ \alpha \over \sigma(t)\,(1-\alpha)})\,, 
\nonumber \\
& & \qquad \qquad \qquad \hspace{2cm}  0 <  \alpha \leq 1,\qquad 
0 < t_a \leq t \leq t_b  < \infty \,, \label{h}
\eea  
where $\sigma(t)$ is a smooth monotonically increasing positive function on 
$[t_a,t_b]$, e.g. $\sigma(t) = t$.

The interpolating partition function on $\Lambda^{(m)}$ constructed 
from $\tilde{c_j}$ and $\tilde{F}_0$ is now defined by    
\beq
\tilde{Z}_{\Lambda^{(m)}}(\beta,h,\alpha,t,r)
= \tilde{F}_0(m,h,\alpha,t)^{|\Lambda^{(m)}|}\,
Z_{\Lambda^{(m)}}(\{\tilde{c}_j(m,\alpha,r)\}) \label{interPF1}
\eeq
where 
\bea
Z_{\Lambda^{(m)}}(\{\tilde{c}_j(m,\alpha,r)\}) 
   & = & 
\int dU_{\Lambda^{(m)}}\;\prod_{p\in \Lambda^{(m)}}\,\Big[ 1 
+ \sum_{j\not= 0} d_j\, \tilde{c}_j(m,\alpha, r)
\, \chi_j(U_p) \Big]  \nonumber \\
 & \equiv & \int dU_{\Lambda^{(m)}}\;\prod_{p\in \Lambda^{(m)}}\, 
f_p(U_p,m,\alpha,r) \,. \label{interPF2}
\eea 
Combining II.1, (\ref{interc3}) and the fact that 
$\tilde{F}_0$ is, by definition, also an increasing function of 
$\alpha$ one has \\
\prop{ 
The interpolating free energies  
$\ln Z_{\Lambda^{(m)}}(\{\tilde{c}_j(m,\alpha,
r)\})$ and $\ln \tilde{Z}_{\Lambda^{(m)}}(\beta,h, 
\alpha,t,r)$ are increasing functions of $\alpha$: 
\beq 
\partial \ln Z_{\Lambda^{(m)}}\Big(\{\tilde{c}_j(m,\alpha,r)
\}\Big) /\partial \alpha \,>\,  0 
\,.  \label{interPFder1}
\eeq
\\
}
Equality in (\ref{interPFder1}) applies only in the 
trivial case were all the coefficients $\tilde{c}_j$'s vanish. 

In terms of (\ref{interPF1}), III.1 and III.2 give 
\beq 
\tilde{Z}_{\Lambda^{(m)}}(\beta,h,0,t,r) 
\leq  Z_{\Lambda^{(m-1)}} \leq 
\tilde{Z}_{\Lambda^{(m)}}(\beta,h,1,t,r) \,. \label{interI1}
\eeq
Now $\tilde{Z}_{\Lambda^{(m)}}(\beta,h,
\alpha,t,r)$ is continuous in $\alpha$. It follows from 
(\ref{interI1}) that there exist a value of 
$\alpha$ in $(0,1)$: 
\beq
\alpha(m, h,t,r,\{c_j(m-1)\},b,\Lambda)\equiv 
\alpha_{\Lambda,h}^{(m)}(t,r)   \label{interI2}
\eeq
such that 
\beq
\tilde{Z}_{\Lambda^{(m)}}(\beta,h,\alpha_{\Lambda,h}^{(m)}(t,r),
t,r) =  Z_{\Lambda^{(m-1)}}  \,.\label{interI3}
\eeq
In other words, at each given value of $t$, $r$, there exist 
a value of $\alpha$ at which the partition 
function on $\Lambda^{(m)}$, resulting from a decimation transformation 
$\Lambda^{(m-1)} \to \Lambda^{(m)}$, equals the partition function on 
$\Lambda^{(m-1)}$. This value is unique by III.3. 
By construction, $\alpha_{\Lambda,h}^{(m)}(t,r)$ is such 
that (\ref{interI3}) remains invariant under variation of 
$t$, $r$ in their domain of definition, i.e.     
$\alpha_{\Lambda,h}^{(m)}(t,r)$ represents the 
level surface of the function $\tilde{Z}_{\Lambda^{(m)}}(\beta,h,\alpha,
t,r)$ fixed by the value $Z_{\Lambda^{(m-1)}}$. 
The parametrization invariance under varying $t$ will be important  
later.

We now examine the dependence   
on $t$, $r$  in (\ref{interI2})  
more closely. Given $Z_{\Lambda^{(m-1)}}$ and 
some interpolation $h$, assume that (\ref{interI3}) is satisfied 
at the point $(t_0, r_0, \alpha=\alpha_{\Lambda,h}^{(m)})$. Then, by the 
implicit function theorem, applicable by III.3, there is a  
function $\alpha_{\Lambda,h}^{(m)}(t,r)$ with 
continuous derivatives such that 
$\alpha_{\Lambda,h}^{(m)}(t_0,r_0)=\alpha_{\Lambda,h}^{(m)}$,  
and uniquely satisfies (\ref{interI3}) in  a sufficiently small neighborhood 
of $ (t_0, r_0, \alpha_{\Lambda,h}^{(m)})$. But since a 
solution to (\ref{interI3}) exists for each choice of $t,r$ in their 
domain of definition, this neighborhood 
can be extended by a standard 
continuity argument to all points of this domain. 
$\alpha_{\Lambda,h}^{(m)}(t,r)$ then represents the regular 
level surface of the function (\ref{interPF1}) fixed  by (\ref{interI3}). 
Furthermore, 
\beq 
{\partial \alpha_{\Lambda,h}^{(m)}(t,r)\over \partial t} = 
v(\alpha_{\Lambda,h}^{(m)}(t,r), t, r) \;, 
\label{alphtder1}
\eeq
where
\beq 
v(\alpha, t, r) \equiv 
- { \D {\partial h(\alpha,t) / \partial t} 
\over{\D  {\partial h(\alpha,t)\over \partial \alpha} + 
A_{\Lambda^{(m)}}(\alpha, r)} }
\;,\label{alphtder2}
\eeq 
with 
\beq
A_{\Lambda^{(m)}}(\alpha, r) \equiv  {1\over \ln F_0^{U}(m) }\,
{1\over |\Lambda^{(m)}| }\, 
{\partial \over \partial\alpha }\ln Z_{\Lambda^{(m)}}\,\Big(\{\tilde{c}_j( m, 
\alpha, r)\}\Big) > 0\;. \label{alphtder3}
\eeq
We will always assume that $h$ is chosen such that 
$\partial h/\partial t$ is negative.  This is the case with the examples 
(\ref{h}). Then, from (\ref{alphtder2}), $v>0$ 
on $0 <\alpha < 1$, with $v=0$ at $\alpha=0$ and $\alpha=1$.

It is also useful to equivalently view $\alpha_{\Lambda,h}^{(m)}(t,r)$ as 
the solution to the ODE 
\bea 
d\alpha/dt & =& v(\alpha, t, r) \,,
\qquad \alpha\in (0,1)\;,\label{ODE}\\
\alpha(t_0) & =&  \alpha_{\Lambda,h}^{(m)} > 0\;,  \qquad t_0 \in [t_a,t_b]\;. 
\nonumber 
\eea 
Then standard results of ODE theory imply the existence of a 
unique solution in a neighborhood of $\alpha_{\Lambda,h}^{(m)}>0$, 
which can in fact be extended indefinitely forward for all 
$t\geq t_0$.\footnote{Indeed, v is differentiable on 
$\alpha_{\Lambda,h}^{(m)}\leq \alpha\leq 1$ and 
vanishes at $\alpha=1$.}

A short computation using (\ref{alphtder1}) gives 
\beq
{d h(\alpha_{\Lambda,h}^{(m)}(t,r),t)\over dt} = 
 - {\partial \alpha_{\Lambda,h}^{(m)}(t,r)\over \partial t} \,
A_{\Lambda^{(m)}}(\alpha_{\Lambda,h}^{(m)}(t,r), r) 
\,, \label{htder}
\eeq
as it should for consistency with (\ref{interI3}).
(\ref{htder}) and (\ref{alphtder1}) make apparent what the 
effect of  a parametrization change due 
to a shift in $t$ is.  
Increasing (decreasing) $t$ increases (decreases) the  contribution of 
$\ln Z_{\Lambda^{(m)}}(\{\tilde{c}_j(m,
\alpha_{\Lambda,h}^{(m)}(t,r),r)\})$ while 
decreasing (increasing) by an equal amount the contribution 
from $\ln F_0^{U}(m)\,h\big(\alpha_{\Lambda,h}^{(m)}(t,r), t
\big)\,|\Lambda^{(m)}|$, so that the sum stays 
constant and equal to $\ln Z_{\Lambda^{(m-1)}}$ in accordance with  
(\ref{interI3}).

The derivative w.r.t. $r$ is similarly given by 
(\ref{alphrder1}) - (\ref{alphrder2}) in Appendix B. 
Now, by (III.1), the upper bound in (\ref{interI1}) 
is optimized for $r=1$, which would appear to make consideration 
of other $r$ values unnecessary.  
The reason one may want, however, to vary 
$r$ away from unity is the following.  

The values $\alpha_{\Lambda,h}^{(m)}(t,r)$ lie in the  
interval $(0,1)$. 
Consider the possibility that one finds that 
$\alpha_{\Lambda,h}^{(m)}(t_m,1)$ differs from $1$ only by terms that 
vanish as the lattice size grows. This means that, since 
$v \geq 0$ in (\ref{alphtder1}), 
$\alpha_{\Lambda,h}^{(m)}(t,1)$ 
is, to within such terms, a constant function of $t$ for all $t\geq t_m$. 
For the purposes of the 
argument in the following sections we want to exclude 
this possibility, and ensure that, at least in some neighborhood of 
a chosen $t$ value, the derivative (\ref{alphtder1}) is non-vanishing 
by an amount independent of lattice size.   

We require that  
\beq 
\delta^\prime  < \alpha_{\Lambda,h}^{(m)}(t,r)< 1-\delta \;,
\label{collar} 
\eeq
with  $\delta > 0$, $\delta^\prime > 0$ independent of the lattice size 
$|\Lambda^{(m)}|$. The lower bound requirement is easily shown (Appendix B) 
to be automatically satisfied by combining II.1 and (\ref{interI3}).  
As it is also shown in Appendix B, one may always ensure that the upper 
bound requirement in (\ref{collar}) holds 
by choosing  the decimation parameter $r$ to 
vary, if necessary, away from unity in the domain 
\beq 
1 \geq r \geq 1-\epsilon  \;, \label{rdomain}
\eeq 
where $0 < \epsilon \ll 1$  with $\epsilon$ independent of $|\Lambda^{(m)}|$.

With (\ref{collar}) in place, (\ref{alphtder1}) and (\ref{htder}) 
imply (Appendix B) that 
\beq 
{\D \partial \alpha_{\Lambda,h}^{(m)}\over \partial t} (t,r)
\geq  \eta_1(\delta) > 0 \, , \qquad \qquad 
- {\D d h\over \D dt}(\alpha_{\Lambda,h}^{(m)}(t,r),t)\geq 
\eta_2(\delta)>0  \, , \label{dercollar} 
\eeq 
where $\eta_1$, $\eta_2$ are lattice-size independent.   
Furthermore, if (\ref{collar}) already holds for $r=1$, 
it also holds for any $r$ in (\ref{rdomain}). 
We may as well then simplify matters in the following by setting   the 
parameter $r$ to the value $r=1-\epsilon$  
with some fixed small $\epsilon$. This $\epsilon$ may eventually be taken 
as small as one pleases after 
a sufficiently large number of decimations have been performed.   
This has an obvious meaning in the context of iterating the 
decimation transformation as pointed out in subsection \ref{disc1} below. 
We accordingly simplify notation by dropping explicit reference to 
$r$, except on occasions when a statement is made for general 
$r$ values. Thus we write $\alpha_{\Lambda,\,h}^{(m)} (t) \equiv 
\alpha_{\Lambda,\,h}^{(m)}(t,1-\epsilon)$, $c^U_j(m) 
\equiv c^U_j(m, 1-\epsilon))$, etc.

\subsection{Representation of the partition function on decimated lattices}
\label{repZ}
So, starting on the original lattice spacing $a$, with partition function 
given in terms of coefficients $\{c_j(\beta)\}$, one may iterate the 
procedure represented by (\ref{interI1}) - (\ref{interI3}). 
Taking the same interpolation family $h$ in every cycle, 
an iteration cycle consists of the following steps.  
\begin{enumerate}
\item[(i)] A decimation transformation $\Lambda^{(m-1)}\to \Lambda^{(m)}$   
given by the rules (\ref{RG1}) - 
(\ref{RG5}) applied to the coefficients in $Z_{\Lambda^{(m-1)}}$, and   
resulting into the upper bound coefficients on $\Lambda^{(m)}$
according to (\ref{upperc}) - (\ref{upperF}) and (\ref{U}). Similarly, 
a lower bound 
on $\Lambda^{(m)}$ is obtained according to (\ref{L})  with lower bound 
coefficients given by (\ref{lowerF1}) and (\ref{lowerc2}) or (\ref{lowerc3}). 
\item[(ii)] Interpolation  between the resulting 
upper and lower bound partition functions on $\Lambda^{(m)}$ 
according to (\ref{interc1}), (\ref{interF1}), and 
(\ref{interPF1}), (\ref{interPF2}). 
\item[(iii)] Fixing the value 
$0 < \alpha_{\Lambda,h}^{(m)}(t) < 1$, eq. (\ref{interI2}), 
so that the $(m-1)$-th step partition function $Z_{\Lambda^{(m-1)}}$ is 
preserved, eq. (\ref{interI3}).  
\item[(iv)] Picking a value of the  parameter $t=t_m$, to fix 
the coefficients 
$\{\tilde{c}_j(m,\alpha_\Lambda^{(m)}(t_m))\}$ of the resulting 
partition function $Z_{\Lambda^{(m)}}$, and return to step (i). 
\end{enumerate}

This scheme for the coefficients in $Z_{\Lambda^{(m)}}$ may be 
depicted as follows:  
\beq 
\begin{array}{c}
\begin{array}{ccc}
\hfill &  c_j(\beta) &  \hfill\\
&  \begin{picture}(60,15)
\put(60,15){\vector(-4,-1){80}}
\end{picture}   
\begin{picture}(38,8)
\put(20,6){\vector(0,-2){20}}
\end{picture}                   
\begin{picture}(60,15)
\put(1,15){\vector(4,-1){80}}
\end{picture}  & \\
 & & \\ 
\end{array} \\
\begin{array}{ccccc}
\hfill \{c_j^L(1)\} & \leq & \{\tilde{c}_j(1,\alpha_{\Lambda,\,h}^{(1)}
(t_1))\} &\leq & \{c_j^U(1)\}\hfill\\
&  \begin{picture}(30,10)
\put(30,10){\vector(-3,-1){50}}
\end{picture}    &   
\begin{picture}(38,8)
\put(20,6){\vector(0,-2){20}}
\end{picture}     &                 
\begin{picture}(30,10)
\put(1,10){\vector(3,-1){50}}
\end{picture}  &  \\
  &  &    &  &   \\
\hfill \{c_j^L(2)\} & \leq & \{\tilde{c}_j(2,\alpha_{\Lambda,\,h}^{(2)}
(t_2))\} &\leq & \{c_j^U(2)\}\hfill\\ 
&  \begin{picture}(30,10)
\put(30,10){\vector(-3,-1){50}}
\end{picture}    &   \begin{picture}(38,8)
\put(20,6){\vector(0,-2){20}}
\end{picture}                 &
\begin{picture}(30,10)
\put(1,10){\vector(3,-1){50}}
\end{picture}  &  \\
&  &   &   & \\
\vdots & & \vdots & & \vdots 
\end{array}\\
\end{array}    \label{S1}
\eeq

The result after $n$ iterations is then: 
\bea
Z_\Lambda(\beta) &= & \int dU_\Lambda\;\prod_{p\in \Lambda}\,f_p(U,\beta) 
\label{O}\\
  & =& \left[\,\prod_{m=1}^n \tilde{F}_0(m,h,\alpha_{\Lambda,\,h}^{(m)}(t_m), 
t_m)^{|\Lambda|/ b^{md}}\,\right]\;
\; Z_{\Lambda^{(n)}}\,\Big(\{\tilde{c}_j( n, \alpha_{\Lambda,\,h}^{(n)}(t_n))
\}\Big) \,. \label{A} 
\eea
(\ref{A}) is an {\it exact integral representation} on the 
decimated lattice $\Lambda^{(n)}$ of the 
partition function $Z_\Lambda$ originally defined on the undecimated 
lattice $\Lambda$ by the integral representation  
(\ref{PF1}) or (\ref{O}).

III.3 allows the iterative procedure leading to (\ref{A}) to be 
implemented in a slightly  different manner, one that turns out later to 
be more convenient for our purposes. Since by III.3  
\beq
Z_{\Lambda^{(m)}}\,\Big(\{\tilde{c}_j(m, \alpha_{\Lambda,\,h}^{(m)}(t_m))\}
\Big) \leq 
Z_{\Lambda^{(m)}}\,\Big(\{\tilde{c}_j(m, 1)\}\Big) = 
Z_{\Lambda^{(m)}}\,\Big(\{c_j^U(m)\}\Big) \,,\label{UPF}
\eeq
an upper bound for each successive iteration step is also obtained by applying 
III.1 to the r.h.s. rather than the l.h.s. of the inequality sign in 
(\ref{UPF}). The only resulting modification  in the above procedure 
is in step (i): the upper bound coefficients  
$c^U_j(m)$ and $F_0^U(m)$ on $\Lambda^{(m)}$  
are computed according to (\ref{upperc}) and (\ref{upperF}) but now 
using the set $\{c_j^U(m-1)\}$ rather than the set 
$\{\tilde{c}_j(m-1, \alpha_{\Lambda,\,h}^{(m-1)}(t_{m-1}))\}$ as the 
coefficient set of the previous step.

The same alternative can be applied to the lower bounds in (\ref{S1}). 
Since, again by III.3, one has
\beq
Z_{\Lambda^{(m)}}\,\Big(\{c_j^L(m)\}\Big) = 
Z_{\Lambda^{(m)}}\,\Big(\{\tilde{c}_j(m, 0)\}\Big) \leq 
Z_{\Lambda^{(m)}}\,\Big(\{\tilde{c}_j(m, \alpha_{\Lambda,\,h}^{(m)}(t_m))\}
\Big) \,, \label{LPF}
\eeq 
a lower bound for each successive iteration step is also obtained by applying 
III.2 to the l.h.s. rather than the r.h.s. of the inequality sign 
in (\ref{LPF}). If one adopts (\ref{lowerc3}), this makes no 
difference since the lower bound coefficients equal zero at every step. 
If one uses (\ref{lowerc2}), the resulting modification to (\ref{S1}) is 
that in step (i) the lower bound coefficients  
$c^L_j(m)$ on $\Lambda^{(m)}$  
are now computed  
using the set $\{c^L_j(m-1)\}$ rather than 
$\{\tilde{c}_j(m-1, \alpha_{\Lambda,\,h}^{(m-1)}(t_{m-1}))\}$ as the 
coefficient set of the previous step.    

One may adopt either or both modifications following from (\ref{UPF}) or 
(\ref{LPF}). Adopting both, the iterative scheme for the coefficients in 
$Z_{\Lambda^{(m)}}$ replacing (\ref{S1}) is:  
\beq 
\begin{array}{c}
\begin{array}{ccc}
\hfill &  c_j(\beta) &  \hfill\\
&  \begin{picture}(60,15)
\put(60,15){\vector(-4,-1){80}}
\end{picture} 
\begin{picture}(20,10)
\put(12,8){\vector(0,-2){20}}
\end{picture}                       
\begin{picture}(60,15)
\put(1,15){\vector(4,-1){80}}
\end{picture}  &  \\
  &  &  \\
\end{array} \\
\begin{array}{ccccc}
\hfill \{ c_j^L(1) \} \qquad & \leq & 
\{\tilde{c}_j(1,\alpha_\Lambda^{(1)}(t_1))\} 
&\leq &  \qquad \{c_j^U(1)\}\hfill\\
 &  &    &  &   \\
\begin{picture}(30,10)
\put(6,10){\vector(0,-2){20}}
\end{picture} &   &   
\begin{picture}(30,10)
\put(18,10){\vector(0,-2){20}}
\end{picture}                &  & 
\qquad \begin{picture}(20,10)
\put(1,10){\vector(0,-2){20}}
\end{picture}  \\
 &  &      &  &    \\
\hfill \{ c_j^L(2) \} \qquad & \leq & 
\{\tilde{c}_j(2,\alpha_\Lambda^{(2)}(t_2))\}
 &\leq & \qquad \{c_j^U(2)\}\hfill  \\
  &   &        &   &     \\
\begin{picture}(30,10)
\put(6,10){\vector(0,-2){20}}
\end{picture}  &    &   
\begin{picture}(30,10)
\put(18,10){\vector(0,-2){20}}
\end{picture}                         &  & 
\qquad \begin{picture}(20,10)
\put(1,10){\vector(0,-2){20}}
\end{picture} \\
  &  &       &  &   \\
\vdots\quad\;  & & \  \vdots & & \ \vdots 
\end{array}\\
\end{array}    \label{S2}
\eeq

This again leads, after $n$ iterations, to the representation (\ref{A}). 
Note, however, that the actual numerical value of  
$\alpha_{\Lambda,\,h}^{(m)}(t_m)$ in (\ref{A}), fixed at each step 
by requiring (\ref{interI3}), will, in general, be different depending on 
whether scheme (\ref{S1}) or (\ref{S2}) is used for the iteration.   
Also note that the upper bounds $c^U_j(m)$ in (\ref{S2}) are not 
optimal compared to those in (\ref{S1}). The scheme (\ref{S2}), however, 
turns out to be more convenient for our purposes in the following.

\subsection{Discussion of the representation (\ref{A})}\label{disc1}
As indicated by the notation, on any finite lattice, the 
$\alpha_{\Lambda,\,h}^{(m)}$ values possess  
a lattice size dependence. This weak dependence enters as a 
correction that vanishes  inversely with lattice size. 
Indeed, by the standard results on the existence of 
the thermodynamic limit of lattice systems, for a partition function 
$Z_{\Lambda^{(m)}}(\{c_j\})$ of the form (\ref{PF2}) 
on lattice $\Lambda^{(m)}$ 
with torus topology (periodic boundary conditions):
\beq
\ln Z_{\Lambda^{(m)}}(\{c_j\})= |\Lambda^{(m)}|\,\varphi(\{c_j\}) + 
\delta\varphi_{\Lambda^{(m)}}(\{c_j\}) \, ,
\eeq
$\varphi(\{c_j\})$ being the free energy per unit volume in the 
infinite volume limit, and  $\delta\varphi_{\Lambda^{(m)}}(\{c_j\})\leq 
O(\mbox{constant})$.\footnote{That is, there are no 
`surface terms' for torus topology. In fact surface terms arising with   
other, e.g. free, boundary conditions can be precisely defined as the 
difference in the free energies computed with periodic versus such other 
boundary conditions \cite{Fi}.} 
From this and (\ref{interI3}) it is straightforward to show that 
\beq
\alpha_{\Lambda,h}^{(m)}(t,r) = 
 \alpha_h^{(m)}(t,r) + \delta 
\alpha_{\Lambda,h}^{(m)}(t,r)  \label{alphsplit1}
\eeq
with $\delta \alpha_{\Lambda,h}^{(m)}(t,r) \to 0$ as some inverse power 
of lattice size in the large volume limit.   
In fact, we have already established the presence of a lattice-size 
independent contribution in $\alpha_{\Lambda,h}^{(m)}(t,r)$ in an 
alternative manner through (\ref{collar}), 
i.e. the fact that  in (\ref{alphsplit1}) one must have 
\beq 
\alpha_h^{(m)}(t,r) > \delta^\prime \;.\label{alphsplit2}
\eeq
An explicit expression for $\delta^\prime$ is given by (\ref{alphlowerb1}),
(\ref{alphlowerb2}). 

At weak and strong coupling the $\alpha_h^{(m)}(t,r)$  
values may be 
estimated analytically by comparison with the weak and 
strong coupling expansions, respectively.  
In general, starting from (\ref{interI1}), the location of 
$\alpha_{\Lambda,\,h}^{(m)}$ satisfying (\ref{interI3}) may be formulated 
as the fixed point of a contraction mapping.  
This allows in principle its numerical determination, for given values of 
all other parameters, to any desired accuracy. 

For our purposes here, however, the actual numerical values 
of the $\alpha_{\Lambda,\,h}^{(m)}$'s, beyond the fact that they are fixed
between $0$ and $1$, will not be directly relevant.  
The main application of the representation (\ref{A}) in this paper will  
be to relate the behavior of the exact theory to 
that of the easily computable approximate decimations bounding it without  
explicit  knowledge of the actual  $\alpha_{\Lambda,\,h}^{(m)}$ values.  

It is important to be clear about the meaning of (\ref{A}). 
The partition function $Z_\Lambda(\beta)$ is originally 
given by its integral representation (\ref{O}) on 
lattice $\Lambda$ of spacing $a$. (\ref{A}) then gives another integral 
representation of $Z_\Lambda(\beta)$ in terms of an integrand defined 
on the coarser lattice $\Lambda^{(n)}$ of spacing $b^na$ plus 
a total bulk free energy contribution resulting from 
decimating between scales $a$ and $b^na$. The action 
$A_p(U,n,\alpha_{\Lambda,h}^{(n)})$ in $\dZ{n}(\{\tilde{c}_j(n,
\alpha_{\Lambda,h}^{(n)}\})$ is constructed to reproduce 
this one physical quantity, i.e. the free energy $\ln Z_\Lambda(\beta)$, 
nothing more and nothing less. In particular, it is {\it not} implied 
that this action on $\Lambda^{(n)}$ can also be used to exactly 
compute any other observable. For that one would need to attempt  
the previous development from scratch with the corresponding 
operator inserted in the integrand.

Recall that, by (\ref{interc3}), the coefficients $\tilde{c}_j(n,\alpha,r)$'s 
are increasing in $\alpha$, and $\tilde{c}_j(n,1,r) =c_j^U(n,r)$, 
$\;\tilde{c}_j(n,0,r) =c_j^L(n)$:   
\beq 
c_j^L(n) <  \tilde{c}_j(\,n,\alpha_{\Lambda,h}^{(n)}(t)\,) <  
c_j^U(n)\; , \qquad \quad 
\quad 0 < \alpha_{\Lambda,h}^{(n)(t)}(t) <  1 \;.\label{cineq5} 
\eeq 
Thus, 
the coefficients $\tilde{c}_j(\,n,\alpha_{\Lambda,h}^{(n)}(t)\,)$  
in the representation (\ref{A}) are bounded from above by 
$c_j^U(n)$ no matter what the actual values of 
$\alpha_{\Lambda,h}^{(n)}(t)$ are.

When considering the implications of this bound under 
successive decimations the advantage of employing scheme (\ref{S2}), 
rather than (\ref{S1}), becomes clear.  
The coefficients $c_j^U(n)$ on the r.h.s. column 
in (\ref{S2}) are obtained by straightforward iteration of the decimation 
rules (\ref{RG2})-(\ref{RG5}) with $\zeta=b^{d-2}$; i.e. only 
knowledge of the $c_j^U(n-1)$, not of the $\tilde{c}_j(n-1,
\alpha_{\Lambda,h}^{(n-1)}(t_{n-1}))$, is 
required to obtain  the $c_j^U(n)$ at the $n$-th step.  
The flow of these $c_j^U(n)$ coefficients then constrains the flow of the 
exact representation coefficients  $\tilde{c}_j(n,
\alpha_{\Lambda,h}^{(n)}(t_n))$ 
according to (\ref{cineq5}) from above. 
In particular, 
{\it if the $c_j^U(n)$'s on the r.h.s. column in (\ref{S2}) 
approach the strong coupling  fixed point, i.e. 
\beq
F_0^U(n)\to 1, \quad \quad  
c_j^U(n) \to 0, \qquad  \mbox{as}\quad  n\to \infty \,,\label{scfp}
\eeq
so must 
the $\tilde{c}_j(n,\alpha_{\Lambda,h}^{(n)})$'s in the representation 
(\ref{A}).}\footnote{
To strictly draw the same conclusion from the alternative scheme 
(\ref{S1}) requires an  
additional step, such as showing that the $c_J^U(n)$'s computed 
according to the scheme (\ref{S1}) flow to the strong coupling regime 
if those computed according to (\ref{S2}) do.}

Now the coefficients $c_j^U(n,r)$ at $r=1$ are the MK decimation
coefficients (cf. section \ref{DEC2}). 
As it is well-known, the MK decimations 
for $SU(2)$ (and also $SU(3)$) are found by explicit evaluation to 
indeed flow to the strong coupling fixed point (\ref{scfp})  
for all starting $\beta<\infty$ and $d\leq 4$. Above the critical dimension 
$d=4$, the decimations result in free spin wave behavior ($c_j^U(n,1) 
\to 1$ as $n\to \infty$) starting from any 
$\beta > \beta_0$, where $\beta_0 =O(1)$.  

Here, for reasons discussed at the end of section \ref{interbounds}, 
we take $r$ in the range (\ref{rdomain}). This may be viewed as 
fixing the direction from 
which the point $\zeta=b^{(d-2)}$, $r=1$ in the parameter space of the 
iteration (\ref{RG2}) - (\ref{RG5}) is approached.  
This is actually irrelevant for the flow behavior of the $ 
c_j^U(n,1-\epsilon)\equiv c_j^U(n)$ since, in the case of $SU(2)$ considered 
here, this point  is a structurally stable point of the 
iteration.\footnote{It is,
however, very much relevant in cases where this point 
is not structurally stable, e.g. in $U(1)$.} 
 
Note that  zero lattice coupling, $g=0$, is a fixed 
point as it is 
for the MK decimations. This is also 
evident from $\lim_{\beta\to\infty}c_j(\beta)=1$ and III.2.

What does (\ref{scfp}) combined with (\ref{cineq5}) 
imply about the question of confinement in the exact theory? 
The 
fact that the long distance part, $\dZ{n}(\{\tilde{c}_j(n,
\alpha_{\Lambda,h}^{(n)})\})$, in (\ref{A}) flows in 
the strong coupling regime does not suffice to answer the question. 
It is the combined contributions 
from all scales between $a$ and $b^na$ in (\ref{A}) that add up to give
the exact free energy $\ln Z_\Lambda(\beta)$. Indeed, recall that,  
by a parametrization change by shifts in $t$ at each decimation step, 
one can shift the relative amounts assigned to these  
various contributions keeping the total sum fixed (cf. remarks immediately 
following (\ref{htder}). This parametrization freedom will in fact 
be important in the following. On the other hand, the fact that 
by (\ref{cineq5})  
the flow of $\tilde{c}_j(n,\alpha_{\Lambda,h}^{(n)}(t_n))$ to the strong 
coupling regime is independent of such parametrization changes 
is strongly suggestive. At any rate, 
to unambiguously determine the long distance behavior of the theory 
one needs to consider appropriate long distance order parameters.

\section{`Twisted' partition function} \label{TZ}
\setcounter{equation}{0}
\setcounter{Roman}{0}
The above derivation leading to the representation (\ref{A}) for 
the partition function  cannot be applied in the presence of 
observables without modification. Thus, in the presence of 
operators involving external sources, such as the Wilson or 
't Hooft loop, translation invariance is lost. Reflection 
positivity is also reduced to hold only in the plane bisecting 
a rectangular loop. 
Fortunately, there are other order parameters that can  
characterize the possible phases of the theory while  
avoiding most of these complications. 
They are the well-known vortex 
free energy, and its transform with respect to the center of 
the gauge group (electric flux free energy).   
They are in fact the natural order parameters in the present context 
since they are constructed out of partition functions, i.e.  
partition functions in the presence of external fluxes. 

Let $Z_\Lambda(\tau_{\mu\nu}, \beta)$ denote the partition function  
with action modified by the `twist' $\tau_{\mu\nu}$, i.e. an element of the 
group center, for every plaquette on a 
coclosed set of plaquettes $\V_{\mu\nu}$ winding through the 
periodic lattice in the $(d-2)$ directions perpendicular to the 
$\mu$, and $\nu$-directions, i.e. winding through every 
$[\mu\nu]$-plane for fixed $\mu, \nu$:
\beq 
A_p(U_p) \to A_p(\tau_{\mu\nu} U_p) \;, \qquad \mbox{if} \quad 
p\in \V_{\mu\nu}\;. \label{twist1}
\eeq  
A nontrivial twist  ($\tau_{\mu\nu}\not=1$) represents a discontinuous 
gauge transformation on the set $\V_{\mu\nu}$ with 
multivaluedness in the group center. Thus, for group $SU(N)$, it introduces 
vortex flux characterized by elements of $\pi_1(SU(N)/Z(N))=Z(N)$. 
The vortex is rendered topologically 
stable by being wrapped around the lattice torus.

In the case of $SU(2)$ explicitly  considered here, there is only one 
nontrivial element, $\tau_{\mu\nu}=-1$.    
As indicated by the notation 
$Z_\Lambda(\tau_{\mu\nu},\beta)$, the twisted partition function  depends only 
on the directions in which $\V_{\mu\nu}$ 
winds through the lattice, not 
the exact shape or location of $\V_{\mu\nu}$. 
This expresses the mod 2 conservation of flux. 
Indeed, a  twist $\tau_{\mu\nu}=-1$ on the plaquettes forming 
a  coclosed set $\V_{\mu\nu}$ can be moved to the plaquettes forming any 
other homologous 
coclosed set $\V^{\ \prime}$ by the change of variables 
$U_b \to -U_b$ for each bond $b$ 
in a set of bonds cobounded by $\V \cup \V^{\ \prime}$, 
leaving  $Z_\Lambda(\tau_{\mu\nu},\beta)$ invariant. By the same token,  
$Z_\Lambda(\tau_{\mu\nu},\beta)$ is invariant under changes 
mod 2 in the number of 
homologous coclosed sets in $\Lambda$ carrying a twist. 
In the following, for definiteness, we fix, say, $\mu=1$, $\nu=2$, and drop  
further explicit reference to the $\mu$, $\nu$ indices. Also, 
we write $Z_\Lambda(-1,\beta) \equiv Z_\Lambda^{(-)}(\beta)$.

(\ref{twist1}) implies that $Z_\Lambda^{(-)}$ is obtained from 
$Z_\Lambda$ by the replacement 
\beq 
f_p(U_p,a) \to f_p(-U_p,a)=
\Big[\, 1 + \sum_{j\not= 0} (-1)^{2j}\,d_j\,c_j(\beta)\,
\chi_j(U_p)\,\Big]  \label{twist2}\,, \qquad \mbox{for each} \quad p\in \V \,,
\eeq
in (\ref{PF1}), (\ref{nexp}), i.e. only half-integer representations on 
plaquettes in $\V$ are 
affected. In general then, the twisted version of the partition function 
(\ref{PF2}) on $\Lambda^{(n)}$ is 
\beq 
Z^{(-)}_{\Lambda^{(n)}}(\{c_j(n)\}) = 
\int dU_{\Lambda^{(n)}}\;
\prod_{p\in \Lambda^{(n)}}\,f^{(-)}_p(U_p,n)  \; , 
\label{PF1atwist}
\eeq 
with  
\beq
f^{(-)}_p(U_p,n) = 
\Big[\, 1 + \sum_{j\not= 0} (-1)^{2j\,S_p[\V]}\,d_j\,c_j(n)\,
\chi_j(U_p)\,\Big]\;.\label{PF1btwist} 
\eeq
$S_p[\V]$ denotes the characteristic function of the plaquette set $\V$, 
i.e. $S_p[\V]=1$ if $p\in \V$, and $S_p[\V]=0$ otherwise. 
A simple result (Appendix A) of obvious physical significance is:  

\prop{ With $c_j(n) \geq 0$, all $j$,  
\beq 
Z^{(-)}_{\Lambda^{(n)}}(\{c_j(n)\}) \leq Z_{\Lambda^{(n)}}(\{c_j(n)\}) \,. 
\label{Z>Z-}
\eeq }
Strict inequality holds in fact in 
(\ref{Z>Z-}) for any nonvanishing $\beta$ on any finite lattice.

Application of the decimation operation 
defined in section \ref{DEC}  on some given 
$Z^{(-)}_{\Lambda^{(m-1)}}$ of the form (\ref{PF1atwist}) results in the rule 
\beq
f^{(-)}_p(U, m-1) \to F_0(m)\,f^{(-)}_p(U,m) = F_0(m) \,
\Big[ 1 + \sum_{j\not= 0} (-1)^{2j\,S_p[\V]}\,d_j\,c_j(m)\,\chi_j(U) \Big]  
\, ,\label{RG1twist} 
\eeq
with coefficients $F_0(m)$, $c_j(m)$ computed according to the rules 
(\ref{RG2}) - (\ref{RG5}). Starting on lattice $\Lambda$, the twisted 
partition function resulting after $n$ such steps is  
\beq 
Z_\Lambda^{(-)}(\beta, n) = \prod_{m=1}^n 
F_0(m)^{|\Lambda|/b^{md}}\; Z_{\Lambda^{(n)}}^{(-)}(\{c_j(n)\})
\; . \label{PF2twist}
\eeq
Note that the flux is carried entirely in 
$Z_{\Lambda^{(n)}}^{(-)}$. Indeed, bulk free energy contributions 
from each $\Lambda^{(m-1)} \to \Lambda^{(m)}$ decimation step arise from local 
moving-integration operations within cells of side 
length $b$ on $\Lambda^{(m-1)}$,   
i.e. topologically trivial subsets, and are thus insensitive to the flux 
presence.  
The evolution with $n$ of the effective action in $Z_{\Lambda^{(n)}}^{(-)}$ 
then determines the manner in which flux spreads, 
which is characteristic of the phase the system is in. 

\subsection{Upper and lower bounds} 
In the presence of the flux, the measure in (\ref{PF1atwist}) possesses 
the property of reflection positivity only in $(d-1)$-dimensional 
planes perpendicular to any one of 
the directions $\rho \not= 1,2$ in which $\V$ winds around the lattice. 
One way of dealing with this is to simply consider the quantity 
\beq
Z^+_{\Lambda^{(n)}}(\{c_j(n)\})\equiv  {1\over 2}
\Big(Z_{\Lambda^{(n)}}(\{c_j(n)\}) + 
Z_{\Lambda^{(n)}}^{(-)}(\{c_j(n)\})\Big) \label{Zplus}
\eeq 
instead of $Z_{\Lambda^{(n)}}^{(-)}$. 
It is indeed easily checked that reflection positivity holds for 
the measure in $Z^+_{\Lambda^{(n)}}$ in all planes. 
A direct consequence of this (Appendix A) is then the analog of II.1:

\prop{For $\dZ{n}^+(\{c_j(n)\})$ given by (\ref{Zplus}) 
with $c_j(n) \geq 0$ for all $j$, and periodic boundary conditions, 

(i) $\dZ{n}^+(\{c_j(n)\})$ is an increasing function of each 
$c_j(m)$: 
\beq 
\partial \dZ{n}^+(\{c_i(n)\}) / \partial c_j(n) \geq 0 \; ;\label{PFplusder0}
\eeq

(ii)
\beq 
\dZ{n}^+(\{c_j(n)\}) \geq \Big[\,1 + \sum_{j\not=0} d_j^2\,
c_j(n)^6 \,\Big]^{|\Lambda^{(n)}|} \; .\label{PFpluslowerb1}
\eeq  
} 
Again, in these bounds equality holds only in the trivial case where 
all $c_j(n)$'s vanish. In particular, one has 
\beq  
\dZ{n}^+(\{c_j(n)\}) \; > \; 1 \,.\label{PFpluslowerb2}
\eeq
Note that these bounds are identical to 
those in II.1. This signifies the obvious fact that they bound from below by 
underestimating the bulk free energies proportional to the lattice volume, 
whereas the lattice size dependence of the free energy discrepancy between 
$\dZ{n}(\{c_j(n)\})$ and $\dZ{n}^{(-)}(\{c_j(n)\})$ is much weaker.

Upper and lower bound statements analogous to III.1 and III.2 can be 
obtained for $Z^+_{\Lambda^{(n)}}$. One has:

\prop{ 
For $Z_{\Lambda^{(n-1)}}^+$ of the form 
(\ref{Zplus}),  
a decimation transformation (\ref{RG1twist}), (\ref{RG2}) - 
(\ref{RG5}) with $\zeta=b^{d-2}$ and $0 < r\leq 1$  results in an upper 
bound on $Z_{\Lambda^{(n-1)}}^+$: 
\beq
Z^+_{\Lambda^{(n-1)}}(\{c_j(n-1)\})\,  \leq \,  
F_0^U(n)^{|\Lambda^{(n)}|}\, 
Z^+_{\Lambda^{(n)}}(\{c_j^U(n,r)\})\;. \label{Uplus}
\eeq 
The r.h.s. in (\ref{Uplus}) is a monotonically 
decreasing function of $r$ on $0 < r\leq 1$.  
}

\prop{ 
For $Z_{\Lambda^{(n-1)}}^+$ of the form (\ref{Zplus}):
\beq 
Z^+_{\Lambda^{(n)}}(\{c_j^L(n)\}) \, \leq \, Z^+_{\Lambda^{(n-1}}(\{ 
c_j(n-1)\}) \;, \label{Lplus}
\eeq 
where the coefficients $c_j^L(n)$ are given by (\ref{lowerc1}).  
}   

The proof of IV.3, as well as that of IV.4, an easy corollary 
of IV.2, are given in Appendix A. It then follows from (\ref{PFplusder0})) 
that (\ref{Lplus}) holds also with coefficients $c_j^L(n)$ given by 
(\ref{lowerc2}) or (\ref{lowerc3}). 
Again, in analogy to III.2, IV.4 also holds 
with $c_j^L$ given by (\ref{lowerc4}), but this form will not be used here.

\subsection{Representation of $Z_\Lambda + Z_\Lambda^{(-)}$ 
on decimated lattices}

The procedure of section \ref{Z}  
leading to the representation (\ref{A}) for $Z_\Lambda$   
can now be applied to $Z^+_\Lambda= (Z_\Lambda + Z_\Lambda^{(-)})/2$.  
One introduces the 
interpolating coefficients $\tilde{c}_j(m,\alpha,r)$
given by eq. (\ref{interc1}), and 
$\tilde{F}_0(m,h,\alpha,t)$ given by eq. (\ref{interF1}) for some 
choice of interpolation function $h$ such as given by the examples 
(\ref{h}). The quantity corresponding to 
(\ref{interPF1}) is then given by  
\beq
\tilde{Z}^+_{\Lambda^{(m)}}(\beta,h,\alpha,t,r)
= \tilde{F}_0(m,h,\alpha,t)^{|\Lambda^{(m)}|}\,
Z^+_{\Lambda^{(m)}}(\{\tilde{c}_j(m,\alpha,r)\}) \label{interPF1plus}
\eeq
where 
\beq
Z^+_{\Lambda^{(m)}}(\{\tilde{c}_j(m,\alpha,r)\}) 
= {1\over2} \Big( Z_{\Lambda^{(m)}}(\{\tilde{c}_j(m,\alpha,r)\}) + 
Z^{(-)}_{\Lambda^{(m)}}(\{\tilde{c}_j(m,\alpha,r)\}) \Big)
\label{interPF2plus}
\eeq
with $Z_{\Lambda^{(m)}}(\{\tilde{c}_j(m,\alpha,r)\})$ given by 
(\ref{interPF2}) and $Z^{(-)}_{\Lambda^{(m)}}(\{\tilde{c}_j(m,\alpha,r)\})$ 
given by (\ref{PF1atwist}) - (\ref{PF1btwist}) with 
coefficients $\tilde{c}_j(m,\alpha,r)$. 
We then have the analog of III.3: 

\prop{ 
The interpolating free energies 
$\ln Z^+_{\Lambda^{(m)}}(\{\tilde{c}_j(m,\alpha,
r)\})$ and 
$\ln \tilde{Z}^+_{\Lambda^{(m)}}(\beta,h,
\alpha,t,r)$ are 
increasing functions of $\alpha$: 
\beq 
\partial \ln Z^+_{\Lambda^{(m)}}\Big(\{\tilde{c}_j(m,\alpha, r)\} \Big) 
/\partial \alpha \, > \,  0 
\,.  \label{interPFder1plus}
\eeq
}

In terms of (\ref{interPF1plus}), IV.3 and IV.4 give 
\beq 
\tilde{Z}^+_{\Lambda^{(m)}}(\beta,h,0,t,r) 
\leq Z^+_{\Lambda^{(m-1)}} \leq 
\tilde{Z}^+_{\Lambda^{(m)}}(\beta,h,1,t,r) \, . \label{interI1plus}
\eeq
which implies that there exist a value of 
$\alpha$ in $(0,1)$:
\beq
\alpha^+(m, h,t,r,\{c_j(m-1)\},b,\Lambda)\equiv 
\alpha_{\Lambda,\, h}^{+(m)}(t,r)   \label{interI2plus}
\eeq
such that 
\beq
\tilde{Z}^+_{\Lambda^{(m)}}(\beta,h,\alpha_{\Lambda,\, h}^{+(m)}(t,r), 
t,r) =  Z^+_{\Lambda^{(m-1)}}  \,.\label{interI3plus}
\eeq
This value is unique, for given values of $t, r$, by IV.5. 
$\alpha_{\Lambda,\, h}^{+(m)}(t,r)$ gives the regular level surface 
of the function $\tilde{Z}^+_{\Lambda^{(m)}}(\beta,h,\alpha,t,r)$ 
fixed by the value $Z^+_{\Lambda^{(m-1)}}$. 

All the considerations concerning the dependence on the parameters 
$t, r$ in the previous section carry over directly to  
$\alpha_{\Lambda,\, h}^{+(m)}(t, r)$. 
In particular, one has 
\beq 
{\partial \alpha_{\Lambda,\, h}^{+(m)}(t,r)\over \partial t} = 
v^+(\alpha_{\Lambda,\,h}^{+(m)}(t,r), t, r) \;, 
\label{alphplustder1}
\eeq
where
\beq 
v^+(\alpha, t, r) \equiv 
- { \D {\partial h(\alpha,t) / \partial t} 
\over{\D  {\partial h(\alpha,t)\over \partial \alpha} + 
A^+_{\Lambda^{(m)}}(\alpha, r)} }
\;,\label{alphplustder2}
\eeq 
with 
\beq
A^+_{\Lambda^{(m)}}(\alpha, r) \equiv  {1\over \ln F_0^{U}(m) }\,
{1\over |\Lambda^{(m)}| }\, 
{\partial \over \partial\alpha }
\ln Z^+_{\Lambda^{(m)}}\,\Big(\{\tilde{c}_j( n, 
\alpha, r)\}\Big) > 0\;. \label{alphplustder3}
\eeq
Again, we always assume that $h$ is chosen such that 
$\partial h/\partial t$ is negative. Then, from (\ref{alphplustder2}), $v^+>0$ 
on $0 <\alpha < 1$, with $v^+=0$ at $\alpha=0$ and $\alpha=1$.  Also 
\beq
{d h(\alpha_{\Lambda,h}^{+(m)}(t,r),t)\over dt} 
= - {\partial \alpha_{\Lambda,h}^{+(m)}(t,r)\over \partial t} \,
A^+_{\Lambda^{(m)}}(\alpha_{\Lambda,h}^{+(m)}(t,r), r) 
\,. \label{hplustder}
\eeq
The derivative w.r.t. $r$ is similarly given by 
(\ref{alphplusrder1}).  

The values (\ref{interI2plus}) obey 
\beq 
\delta^{+\,\prime} < \alpha_{\Lambda,\,h}^{+(m)}(t,r)< 
1-\delta^+  \label{collarplus} 
\eeq 
with lattice-size independent, positive $\delta^+$ and $\delta^{+\,\prime}$. 
Again, the lower bound is automatically satisfied, whereas the upper 
bound is ensured by letting the parameter $r$ vary, if necessary,  
in (\ref{rdomain}) (cf. Appendix B).  
From this it follows that the analog of (\ref{dercollar}):  
\beq 
{\D \partial \alpha_{\Lambda,\,h}^{+(m)}\over \partial t} (t,r)
\geq  \eta^+_1(\delta^+) > 0 \, , \qquad \qquad 
- {\D d h\over \D dt}(\alpha_{\Lambda,\,h}^{+(m)}(t,r),t) \geq 
\eta^+_2(\delta^+)>0  \, , \label{dercollarplus} 
\eeq 
holds for some lattice-size independent $\eta^+_1$, $\eta^+_2$. 
Since, furthermore, (\ref{collarplus}) holds for any $r$ if it already 
holds for $r=1$, we may again set  
$r = 1-\epsilon$, and, according to the convention introduced in the 
previous section, write $\alpha_{\Lambda,\,h}^{+(m)}(t) \equiv 
\alpha_{\Lambda,\,h}^{+(m)}(t, 1 -\epsilon)$, 
etc.

As in the last section, one may iterate this procedure of 
performing a decimation transformation to produce upper and lower bounds 
according to (\ref{interI1plus}), and then fixing the value 
(\ref{interI2plus}) of the interpolating parameter $\alpha$ according to 
(\ref{interI3plus}). 
Assume that we choose the same interpolation family $h$ at every step.  
Then starting from the original lattice, after $n$ iterations one obtains 
\bea
Z^+_\Lambda(\beta) &= & {1\over 2} \Big( 
Z_\Lambda(\beta) + Z^{(-)}_\Lambda(\beta)\Big) 
\nonumber \\
  & =& \left[\,\prod_{m=1}^n \tilde{F}_0(m, h,\alpha_{\Lambda,\, 
h}^{+(m)}(t_m), t_m)^{|\Lambda|/ b^{md}}\,\right]\;
\; Z_{\Lambda^{(n)}}^+\,\Big(\{\tilde{c}_j( n, \alpha_{\Lambda,\, h}^{+(n)}
(t_n))\}\Big) \,. \label{B} 
\eea

The discussion in subsection \ref{disc1} concerning the representation 
(\ref{A}) of $Z_\Lambda$ applies equally well to (\ref{B}). In particular, 
note that again the existence of the large volume limit implies that 
\beq
\alpha_{\Lambda,h}^{+(m)}(t,r) = 
 \alpha_h^{+(m)}(t,r) + \delta 
\alpha_{\Lambda,h}^{+(m)}(t,r)  \label{alphplussplit1}
\eeq
with $\delta \alpha_{\Lambda,h}^{+(m)}(t,r) \to 0$ as some inverse power of 
lattice size in the  $|\Lambda^{(m)}|\to \infty$ limit. 
Alternatively, (\ref{collarplus}) already implies that one must have 
a lattice-size independent contribution $\alpha_h^{+(m)}(t,r) \,>\, 
\delta^{+\,\prime}\,>\,0$ 
in (\ref{alphplussplit1}) (cf. Appendix B).

Again, either scheme (\ref{S1}) or (\ref{S2}) may be used to 
obtain (\ref{B}). For the reasons already noted, however,   
the latter scheme is more convenient for our considerations. 
Note, furthermore, that the bounding coefficients 
$c^U_j(m)$ and $c_j^L(m)$ in this scheme are 
the same for $Z_\Lambda$ and $Z^+_\Lambda$ since they do not depend 
on $\alpha_{\Lambda,\,h}^{(m-1)}$ or $\alpha_{\Lambda,\,h}^{+(m-1)}$. 
We, therefore, adopt it in what follows as the common iteration scheme 
for  $Z_\Lambda$ and $Z^+_\Lambda$:  
\beq 
\begin{array}{c}
\begin{array}{ccc}
 & \quad   c_j(\beta) \qquad &  \\
&  \begin{picture}(60,20)
\put(45,20){\vector(-4,-1){100}}
\end{picture}   
\begin{picture}(30,20)\put(8,20){\vector(0,-2){20}}\end{picture}      
\begin{picture}(60,20)
\put(1,20){\vector(4,-1){100}}
\end{picture}  &  
\end{array} \\
\begin{array}{ccccc}
\hfill \{c_j^L(1)\} \quad & \leq & 
\{\tilde{c}_j(1,\alpha_{\Lambda,\,h}^{(1)}(t_1))\}, \; 
\{\tilde{c}_j(1,\alpha_{\Lambda,\,h^+}^{+(1)}(t^+_1))\}
&\leq & \quad \{c_j^U(1)\}\hfill\\
\begin{picture}(30,20)
\put(8,20){\vector(0,-2){20}}
\end{picture} &   &   
\begin{picture}(30,20)\put(10,20){\vector(0,-2){20}}\end{picture}      
          &  & 
\qquad \begin{picture}(30,20)
\put(1,20){\vector(0,-2){20}}
\end{picture}  \\
\hfill \{c_j^L(2)\} \quad & \leq & 
\{\tilde{c}_j(2,\alpha_{\Lambda,\,h}^{(2)}(t_2))\}, \; 
\{\tilde{c}_j(2,\alpha_{\Lambda,\,h^+}^{+(2)}(t^+_2))\}
&\leq & \quad \{c_j^U(2)\}\hfill  \\
\begin{picture}(30,20)
\put(8,20){\vector(0,-2){20}}
\end{picture}  &    &     
\begin{picture}(30,20)\put(10,20){\vector(0,-2){20}}\end{picture}  
              &  & 
\qquad \begin{picture}(30,20)
\put(1,20){\vector(0,-2){20}}
\end{picture} \\
\vdots \; \quad & & \vdots & & \vdots \;
\end{array}\\
\end{array}    \label{S3}
\eeq
In (\ref{S3}) and in the following, the more detailed notation $h^+$ and 
$t^+$ is used for the choice of interpolation and 
$t$-parameter values occurring in (\ref{B}) whenever they  need be 
distinguished from those used in the representation (\ref{A}) for 
$Z_\Lambda$, which can, of course, be chosen independently.

As indicated by the notation, even for common choice of interpolation 
$h=h^+$ and of all other parameters, the values of 
$\alpha_{\Lambda,\,h}^{+(m)}(t ,r)$ 
fixed by the requirement (\ref{interI3plus}) are a priori distinct 
from those of $\alpha_{\Lambda,\,h}^{(m)}(t,r)$ fixed by 
(\ref{interI3}). It is easily seen, however, that for sufficiently  
large lattice volume they must nearly coincide. We examine this 
difference more precisely below.

\section{The ratio $Z_\Lambda^{(-)}/Z_\Lambda$}\label{Z-/Z}
\setcounter{equation}{0}
\setcounter{Roman}{0}

We may now compare $Z_\Lambda$ and $Z_\Lambda + Z^{(-)}_\Lambda$ by 
means of their representations (\ref{A}) and (\ref{B}) on 
successively decimated lattices.  
Consider then the ratio of $Z_\Lambda + Z_\Lambda^{(-)}$ and $Z_\Lambda$ as 
given by (\ref{B}) and (\ref{A}) with common choice 
of interpolation $h=h^+$ after one decimation: 
\bea
\left(\,1+ {Z_\Lambda^{(-)} \over Z_\Lambda }\,\right) & = & 
{ 2 \tilde{Z}^+_{\Lambda^{(1)}}\,(\beta, h,  
\alpha_{\Lambda,\, h}^{+(1)}(t^+),\, t^+) \over 
\tilde{Z}_{\Lambda^{(1)}}\,(\beta, h,  
\alpha_{\Lambda,\,h}^{(1)}(t),\, t) } \label{ratio1a} \\
& = & \left(\,{ \tilde{Z}_{\Lambda^{(1)}}\,(\beta, h,  
\alpha_{\Lambda,\, h}^{+(1)}(t^+),\, t^+) \over 
\tilde{Z}_{\Lambda^{(1)}}\,(\beta, h,  
\alpha_{\Lambda,\,h}^{(1)}(t),\, t) }\,\right) \left(\,1+ 
{ Z_{\Lambda^{(1)}}^{(-)}\,\Big(\{\tilde{c}_j(1,\alpha_{\Lambda,\,h}^{+(1)}
(t^+))\}\Big)
\over Z_{\Lambda^{(1)}}\,\Big(\{\tilde{c}_j(1,\alpha_{\Lambda,\,h}^{+(1)}
(t^+))\}\Big) } 
\,\right) 
\label{ratio1}
\eea 
By construction, the r.h.s. is invariant under independent variations 
of $t$ and $t^+$. 
Now since by IV.1 
\beq
 1< \left(\,1+ {Z_\Lambda^{(-)} \over Z_\Lambda }\,\right) < 2 \; \qquad 
\mbox{and} \qquad 
1 < \left(\,1+ 
{ Z_{\Lambda^{(1)}}^{(-)}\,\Big(\{\tilde{c}_j(1,\alpha_{\Lambda,\,h}^{+(1)}
(t^+))\}\Big)
\over Z_{\Lambda^{(1)}}\,\Big(\{\tilde{c}_j(1,\alpha_{\Lambda,\,h}^{+(1)}(t^+))
\}\Big) } 
\,\right) <2 \;,\label{ratiobounds}
\eeq
it follows that\footnote{(\ref{ratioconstr}) clearly 
holds for general values of the parameter $r$, not just for the values 
(\ref{rdomain}) used in (\ref{ratio1}).} 
\beq 
{1\over 2} < { \tilde{Z}_{\Lambda^{(1)}}\,(\beta, h,  
\alpha_{\Lambda,\, h}^{+(1)}(t^+),\, t^+) \over 
\tilde{Z}_{\Lambda^{(1)}}\,(\beta, h,  
\alpha_{\Lambda,\,h}^{(1)}(t),\,  t) }
 < 2 \;. \label{ratioconstr}
\eeq
Though the bounds (\ref{ratiobounds}) are rather crude, the resulting 
constraint (\ref{ratioconstr}) is quite informative. 
First, it says that if in the equality (\ref{interI3}), i.e.  
\[ \tilde{Z}_{\Lambda^{(1)}}\,(\beta, h,  
\alpha_{\Lambda,\,h}^{(1)}(t), t) = Z_\Lambda \] 
one substitutes for $\alpha_{\Lambda,\,h}^{(1)}(t)$ the wrong level 
surface $\alpha_{\Lambda,\, h}^{+(1)}(t)$, 
the resulting discrepancy in the free energy per unit volume 
is at most $O(1/|\Lambda^{(1)}|)$. Furthermore, (\ref{ratioconstr}) 
constrains by how much $\alpha_{\Lambda,h}^{+(1)}(t)$  
can differ from $\alpha_{\Lambda,h}^{+(1)}(t^+)$ at $t=t^+$.  
From the definition (\ref{interPF1}) and III.3, the change  
in $\tilde{Z}_{\Lambda^{(1)}}\,
(\beta,h, \alpha, t, r)$
under a shift $\delta \alpha$ in $\alpha$ satisfies 
\beq
|\,\delta \ln \tilde{Z}_{\Lambda^{(1)}}\,
(\beta, h, \alpha, t, r)| >  
|\,\delta \alpha |\,|\Lambda^{(1)}| \ln F_0^U(1)\, 
{\partial h(\alpha,t)\over \partial \alpha} \;. \label{varcon}
\eeq
When combined 
with (\ref{varcon}), the constraint (\ref{ratioconstr}), taken at general $r$,
implies that one  
must have 
\beq
| \alpha_{\Lambda,\,h}^{+(1)}(t,r) - 
\alpha_{\Lambda,\,h}^{(1)}(t,r) | \leq 
O({1\over |\Lambda^{(1)}|}) \;. \label{alphdiff}
\eeq 
This implies that in (\ref{alphsplit1}), (\ref{alphplussplit1}) one 
has $\alpha_h^{(1)}(t)=\alpha_h^{+(1)}(t)$, i.e. any difference occurs only in 
the parts $\delta \alpha_{\Lambda,\,h}^{(1)}$, $\delta \alpha_{\Lambda,\,h}^{
+(1)}$ that vary inversely with lattice size. Thus, in the large volume 
limit, this difference becomes unimportant if one is 
interested only in the computation of partition functions, or  
bulk free energies. This, however, is 
not the case for free energy differences such as the ratio (\ref{ratio1a}). 
Indeed, any discrepancy of the size (\ref{alphdiff}) means that 
the first factor in (\ref{ratio1}) can contribute as much as the second 
factor in round brackets. Thus the expression  for the ratio 
of the twisted to the untwisted partition function given by (\ref{ratio1}), 
though exact, is not immediately useful for extracting  this 
ratio on the coarser lattice.

To address this issue one may make use of the $t$-parametrization invariance 
of (\ref{ratio1}).   
First  the cancellation of 
the bulk energies generated in the integration from 
scale $a$ to $ba$ is made explicit as follows. 
For any given $t^+_1$, choose $t_1$ in  $\tilde{Z}_{\Lambda^{(1)}}
\,(\beta,  \alpha_{\Lambda,h}^{(1)}(t_1), t_1)$ so 
that 
\beq
h(\alpha_{\Lambda,\,h}^{(1)}(t_1), t_1) = 
h(\alpha_{\Lambda,\,h}^{+(1)}(t^+_1), t^+_1) \;.\label{hequal1}
\eeq
This is clearly always possible by 
(\ref{dercollar}) and (\ref{dercollarplus}), and by (\ref{alphdiff}); 
in fact,  $t_1-t^+_1 = O(1/|\Lambda^{(1)}|)$.  
Then (\ref{ratio1a}) assumes the form 
\beq
\left(\,1+ {Z_\Lambda^{(-)} \over Z_\Lambda }\,\right) = 
{ 2 Z^+_{\Lambda^{(1)}}\,\Big(\{\tilde{c}_j(1,\alpha^+_{\Lambda,\,h}(t^+_1))
\}\Big)
\over Z_{\Lambda^{(1)}}\,\Big(\{\tilde{c}_j(1,\alpha_{\Lambda,\,h}(t_1))
\}\Big) } \;. \label{ratio2}
\eeq

We may now iterate this  procedure  performing $(n-1)$ decimation steps 
according to the scheme (\ref{S3}), at each step 
choosing  $t_m$, $t^+_m$ such that 
\beq
h(\,\alpha_{\Lambda,\,h}^{(m)}(t_m), t_m) = 
h(\,\alpha_{\Lambda,\,h}^{+(m)}(t^+_m), t^+_m) \;, \qquad 
m=1,\ldots (n-1) \;.  \label{hequal2}
\eeq

Carrying out a final $n$-th decimation step one obtains 
\bea
\left(\,1+ {Z_\Lambda^{(-)} \over Z_\Lambda }\,\right) & = & 
{ 2\, \tilde{Z}^+_{\Lambda^{(n)}}\,(\beta, h,  
\alpha_{\Lambda,\, h}^{+(n)}(t^+),\, t^+) \over 
\tilde{Z}_{\Lambda^{(n)}}\,(\beta, h,  
\alpha_{\Lambda,\,h}^{(n)}(t),\, t) } \label{ratio3} \\
 & = & { \tilde{Z}_{\Lambda^{(n)}}\,(\beta, h,  
\alpha_{\Lambda,\,h}^{+(n)}(t^+),\, t^+) \over \tilde{Z}_{\Lambda^{(n)}}\,
(\beta, h,  \alpha_{\Lambda,\,h}^{(n)}(t),\, t)  } \,\left(\,1+ 
{ Z_{\Lambda^{(n)}}^{(-)}\,\Big(\{\,\tilde{c}_j(n,\alpha_{\Lambda,\,h}^{+(n)}
(t^+))\,\}\Big)
\over Z_{\Lambda^{(n)}}\,\Big(\{\,\tilde{c}_j(n,\alpha_{\Lambda,\,h}^{+(n)}
(t^+))\,\}\Big) } \right)
\label{ratio4}
\eea 
The argument for $n=1$ (eq. (\ref{ratio1})) above may now be applied to 
(\ref{ratio4}) to conclude 
\beq 
| \alpha_{\Lambda,h}^{+(n)}(t,r) - 
\alpha_{\Lambda,h}^{(n)}(t,r) | \leq 
O({1\over |\Lambda^{(n)}|}) \;. \label{alphdiffN}
\eeq 
Any such discrepancy between $\alpha_{\Lambda,h}^{+(n)}(t)$ and 
$\alpha_{\Lambda,h}^{(n)}(t)$ in (\ref{ratio4}) presents the same problem 
for extracting the ratio at scale $b^n a$ as at scale $ba$.  
In this  sense  (\ref{ratio4}) is not qualitatively different from 
the $n=1$ case (\ref{ratio1}). Transferring the discrepancy to 
large $n$, however, allows a technical simplification as we see below. 

Next, consider (\ref{ratio3}) rewritten as 
\beq
\left(\,1+ {Z_\Lambda^{(-)} \over Z_\Lambda }\,\right) = 
\left( { Z^+_{\Lambda^{(n-1)}} \over \tilde{Z}^+_{\Lambda^{(n)}}\,(\beta,  h,
\alpha_{\Lambda,\,h}^{(n)}(t),\, t) } \right)\,\left(\,1+ 
{ Z_{\Lambda^{(n)}}^{(-)}\,\Big(\{\,\tilde{c}_j(n,\alpha_{\Lambda,\,h}^{(n)}
(t))\,\}\Big)
\over Z_{\Lambda^{(n)}}\,\Big(\{\, \tilde{c}_j(n,\alpha_{\Lambda,\,h}^{(n)}
(t))\,\}\Big) }
\,\right) 
\label{ratio5}
\eeq  
by use of (\ref{interI3plus}). By construction (cf. (\ref{interI3})), 
$\alpha_{\Lambda,\,h}^{(n)}(t)$ is such 
that the r.h.s. in (\ref{ratio3}), hence in (\ref{ratio5}), 
is invariant under changes in the 
parameter $t$; but note 
that the two $\alpha_{\Lambda,\,h}^{(n)}$-dependent factors in 
round brackets on the r.h.s. in (\ref{ratio5}) are {\it not} 
separately invariant. 
If, for some given $t$, $\alpha_{\Lambda,\,h}^{(n)}(t)$ is larger 
(smaller) than $\alpha_{\Lambda,\,h}^{+(n)}(t)$, then, 
by IV.5, $\tilde{Z}^+_\Lambda\,(\beta, h, 
\alpha_{\Lambda,\,h}^{(n)}(t),\,t)$ is larger (smaller) than 
$\tilde{Z}^+_{\Lambda^{(n)}}\,(\beta,h, \alpha_{\Lambda,\,h}^{+(n)}(t),
\, t)= Z^+_{\Lambda^{(n-1)}}$, 
and the second factor in round brackets on the r.h.s. of (\ref{ratio5}) 
overestimates (underestimates) the ratio $Z_\Lambda^{(-)}/ Z_\Lambda $. 
It is then natural to ask whether there exist a value $t=  
t_{\Lambda,h}^{(n)}$ such that 
\beq  
\tilde{Z}^+_{\Lambda^{(n)}}\,\Big(\beta, h,\alpha_{\Lambda, h}^{(n)}(
t_{\Lambda,h}^{(n)}),\, t_{\Lambda,h}^{(n)} \Big) 
= Z^+_{\Lambda^{(n-1)}} \;.\label{interIfixplus}
\eeq 
Note that 
the graphs of $\alpha_{\Lambda, h}^{(n)}(t)$ and 
$\alpha_{\Lambda, h}^{+(n)}(t)$ must intersect at $t_{\Lambda,\,h}^{(n)}$. 
 
A unique solution to (\ref{interIfixplus}) indeed exists as shown in 
Appendix C provided  
\beq
A_{\Lambda^{(n)}}(\alpha, r)  \geq
A^+_{\Lambda^{(n)}}(\alpha, r)   \label{A>A+}
\eeq
with $r$ in (\ref{rdomain}).  An equivalent statement to 
(\ref{A>A+}) is 
\beq 
A_{\Lambda^{(n)}}(\alpha, r)  \geq
A^{(-)}_{\Lambda^{(n)}}(\alpha, r) \, ,  \label{A>A-}
\eeq
where $A^{(-)}_{\Lambda^{(m)}}(\alpha, r)$ is defined 
by (\ref{alphplustder3}) but with $Z^+_{\Lambda^{(n)}}$ replaced by 
$Z_{\Lambda^{(n)}}^{(-)}$. 
Assume now that under successive decimations the coefficients 
$c^U_j(m)$ in (\ref{S3}) evolve within the convergence radius of the 
strong coupling cluster expansion. Taking then $n$  in 
(\ref{ratio3}) sufficiently 
large, we need establish inequality (\ref{A>A+})\footnote{For Abelian 
systems, comparison inequalities of 
the type (\ref{A>A+}) either follow from Griffith's inequalities, or can be 
approached by the same methods. All such known 
methods fail in the non-Abelian case.} only at 
strong coupling. Within this expansion it is a straightforward exercise 
to establish the validity of (\ref{A>A+}), with strict inequality 
on any finite lattice.

We summarize the above development in the following:\\
\prop{
Consider $n$ successive decimation steps performed according to the 
scheme (\ref{S3}). 
Assume that there is an $n_0$ such that the upper bound coefficients 
$c^U_j(n)$ become sufficiently small for $n\geq n_0$. 
 
Then the ratio of the twisted to the untwisted partition 
function on lattice $\Lambda$, of spacing $a$, has a representation on lattice 
$\Lambda^{(n)}$, of spacing $b^na$ and $n \geq n_0$, given by:
\beq
{Z_\Lambda^{(-)}(\beta) \over Z_\Lambda(\beta) } = 
{ Z_{\Lambda^{(n)}}^{(-)}\,\Big(\{\,\tilde{c}_j(n,\alpha_\Lambda^{*\,(n)})\,\}
\Big) \over Z_{\Lambda^{(n)}}\,\Big(\{\, \tilde{c}_j(n,\alpha_\Lambda^{*\,(n)})
\,\}\Big) } \;,
\label{ratio6}
\eeq 
where 
\beq
\alpha_\Lambda^{*\,(n)}\equiv \alpha_{\Lambda, h}^{(n)}(t_{\Lambda,\,h}^{(n)})
 \;.\label{alphstar}
\eeq
Here, the function $\alpha_{\Lambda, h}^{(n)}(t)$ is defined 
by (\ref{interI3}), i.e. is the solution for $\alpha$ to 
\beq
\tilde{Z}_{\Lambda^{(n)}}\,(\beta, h, \alpha,\, t) = 
Z_{\Lambda^{(n-1)}}\;, \label{interI3A}
\eeq 
and $t_{\Lambda,\,h}^{(n)}$ is defined by 
(\ref{interIfixplus}), i.e. is the solution for $t$ to the equation 
\beq  
\tilde{Z}^+_{\Lambda^{(n)}}\,(\beta, h,\alpha_{\Lambda, h}^{(n)}(t),\,t) 
= Z^+_{\Lambda^{(n-1)}} \;.\label{interIfixplusA}
\eeq 
}

As indicated by the notation in (\ref{alphstar}), 
any dependence on $h$ must cancel in $\alpha_\Lambda^{*\,(n)}$.  
Indeed, Cauchy's form of the intermediate  value theorem gives 
\beq
{ \ln Z_{\Lambda^{(n)}}^{(-)}\,\Big(\{\,\tilde{c}_j(n,\alpha)\,\}\Big) -  
\ln Z_{\Lambda^{(n)}}^{(-)}\,\Big(\{\,\tilde{c}_j(n,\alpha_\Lambda^{*\,(n)})
\,\}\Big)
\over \ln Z_{\Lambda^{(n)}}\,\Big(\{\,\tilde{c}_j(n,\alpha)\,\}\Big) - 
\ln Z_{\Lambda^{(n)}}\,\Big(\{\,\tilde{c}_j(n,\alpha_\Lambda^{*\,(n)})\,\}
\Big) } 
= { A_{\Lambda^{(n)}}^{(-)}(\xi) \over A_{\Lambda^{(n)}}(\xi) }
\leq 1 \;, \label{Cauchy}
\eeq
for some $\xi$ between $\alpha_\Lambda^{*\,(n)}$ and $\alpha$,
and use of (\ref{A>A-}) was made to obtain the last  inequality.  
Setting $\alpha$ equal to $1$ in (\ref{Cauchy}), combining with 
(\ref{ratio6}), and using III.3, IV.5, gives 
\beq 
{Z_\Lambda^{(-)} \over Z_\Lambda } \geq  
{ Z_{\Lambda^{(n)}}^{(-)}\,\Big(\{\,c^U_j(n)\,\}\Big)  
\over Z_{\Lambda^{(n)}}\,\Big(\{\,c^U_j(n)\,\}\Big) } \; . 
\label{ratiolower} 
\eeq 
The upper bound coefficients in (\ref{S3}) then, which correspond to  
upper bounds for the partition functions $Z_\Lambda$ and $Z_\Lambda^{(-)}$,  
give a lower bound for the ratio  
$Z_\Lambda^{(-)}/Z_\Lambda$.\footnote{This result was first 
stated a long time ago in \cite{T}.} 
Setting $\alpha=0$ in (\ref{Cauchy}), similarly yields an upper bound. 
Thus:\\
\prop{
With  the same conditions as in V.1 the 
ratio of the twisted to the untwisted partition 
function on lattice $\Lambda$ of spacing $a$ is bounded on lattice 
$\Lambda^{(n)}$ of spacing $b^na$ by:
\beq    
{ Z_{\Lambda^{(n)}}^{(-)}\,\Big(\{\,c^L_j(n)\,\}\Big)  
\over Z_{\Lambda^{(n)}}\,\Big(\{\,c^L_j(n)\,\}\Big) } 
\geq 
{Z_\Lambda^{(-)} \over Z_\Lambda } 
\geq  
{ Z_{\Lambda^{(n)}}^{(-)}\,\Big(\{\,c^U_j(n)\,\}\Big)  
\over Z_{\Lambda^{(n)}}\,\Big(\{\,c^U_j(n)\,\}\Big) } \; . 
\label{ratiolowerupper}
\eeq 
}

Now, the ratio of the interpolating 
partition functions (\ref{interPF2plus}) 
and (\ref{interPF2}) interpolates monotonically between 
the upper and lower bounds in (\ref{ratiolowerupper}) since 
\beq 
{d\over d\alpha }\; { Z^{(-)}_{\Lambda^{(n)}}(\{\tilde{c}_j(n,\alpha\}) 
\over 
Z_{\Lambda^{(n)}}(\{\tilde{c}_j(n,\alpha)\}) } < 0 \label{ratioder}
\eeq 
by (\ref{A>A-}). It follows that there exist a unique value 
$\alpha^{*\,(n)}_\Lambda$ of $\alpha$ at which this ratio 
of the interpolating partition functions equals 
$Z_\Lambda^{(-)} / Z_\Lambda $. 
This is a restatement of (\ref{ratio6}), but makes explicit the fact that 
this value is independent of $h$. In fact, it shows that all dependence 
on parametrization choices, i.e. the choice of parameters $t_m$ made 
in successive decimations, eventually cancels in  
$\alpha^{*\,(n)}_\Lambda$. Indeed, the latter can depend only on the number 
of decimations $n$ and the initial coupling $\beta$, since this is all the 
upper and lower bounds in (\ref{ratiolowerupper}) depend on. 
This, in retrospect, is as expected, since all bulk free-energy 
contributions depending on such choices were canceled in finally 
arriving at (\ref{ratio6}), but V.2 makes it manifest.

(\ref{ratiolowerupper}) was obtained as a corollary of (\ref{ratio6}).  
An alternative approach would be to proceed in 
the reverse direction, i.e. 
establish (\ref{ratiolowerupper}) 
directly, from which (\ref{ratio6}) would follow by interpolation 
between the upper and lower bounds as in the previous paragraph. 
In other words, follow also in the case of the 
ratio of the partition functions the approach followed separately for 
the untwisted and twisted partition functions in the previous sections. 
This is further discussed in Appendix C.

\section{Confinement}\label{CONf}
\setcounter{equation}{0}
\setcounter{Roman}{0}

\subsection{Order parameters}\label{CONFOP}
The vortex free energy $F_\Lambda^{(-)}$ is defined by  the 
ratio of partition functions considered in the previous section:   
\beq 
\exp(-F_\Lambda^{(-)}(\beta)) = {Z_\Lambda^{(-)}(\beta) 
\over Z_\Lambda(\beta)} \;. 
\label{vfe}
\eeq
It represents the free energy cost for adding a vortex to the vacuum,  
the $Z(2)$ flux of the inserted vortex being rendered stable by 
wrapping  around the toroidal lattice. As has been discussed in the 
literature, all possible phases of gauge theory (Higgs, Coulomb, or 
confinement) can be characterized by the behavior of (\ref{vfe}) 
as one lets the lattice become large. In particular, having 
taken the vortex to wind through the  
lattice in the directions $\kappa=3, \ldots, d$,  
a confining phase is signaled by the asymptotic behavior 
\beq 
F_\Lambda^{(-)}(\beta) \sim L\,
\exp(\,-\hat{\sigma}(\beta) |A|\,) \;, \label{vfeconf}
\eeq
where $L\equiv \prod_{\kappa\not= 1,2}\,L_\kappa$, and 
$A\equiv L_1 L_2$. (\ref{vfeconf}) represents exponential spreading  
of the flux introduced by the twist on the set $\V$ in the 
transverse directions (creation of mass gap), with $ \hat{\sigma}(\beta)$ 
giving the exact string tension. 
Note that, according to (\ref{vfeconf}), $F_\Lambda^{(-)}(\beta))\to 0$ 
as $|\Lambda|\to \infty $ in any power-law fashion, i.e. one 
has `condensation' of the vortex flux.  
The behavior (\ref{vfeconf}) is dictated by physical reasoning \cite{tH}, 
\cite{MP},  
and explicitly realized within the strong coupling expansion. 
As such free energies differences are generally notoriously difficult to 
measure  accurately, demonstration of the behavior (\ref{vfeconf}) 
by numerical simulations at large $\beta$'s has been achieved only 
relatively recently \cite{KT}, \cite{Fetal}.

The $Z(2)$ Fourier transform of (\ref{vfe})
\beq 
\exp(-F_\Lambda^{\rm el}(\beta)) = {1\over 2}\Big( 
\,1 - {Z_\Lambda^{(-)}(\beta) 
\over Z_\Lambda(\beta)} \,\Big) 
\label{efe}
\eeq
gives the corresponding dual (w.r.t. the gauge group center) order 
parameter, the color electric free energy.  (\ref{vfe}) and (\ref{efe}) 
are ideal pure long-range order parameters. They do not suffer from 
the physically irrelevant but technically quite bothersome  
complications, such as loss of translational 
invariance, or mass renormalization and other short 
range contributions, that arise from the explicit introduction of 
external sources. Such external current sources are introduced  
in the definition of the Wilson and t'Hooft loops. Furthermore, the 
behavior of the latter can be bounded by that of 
(\ref{vfe}) and (\ref{efe}) \cite{TY}. 
In particular, the  following relation holds. Let $C$ be a rectangular loop 
of minimal area $S$ lying in a $2$-dimensional $[12]$-plane. Then 
\cite{TY}:
\beq 
\vev{W[C]}_\Lambda \leq \left[ \exp(-F_\Lambda^{\rm el}\right]^{S/A} \,, 
\label{W-vfebound} 
\eeq
where $W[C]=\chi_{_{1/2}}\,\Big(\prod_{b\in C} U_b\Big)$ is the usual 
Wilson loop observable. It follows from (\ref{W-vfebound}) that 
confining behavior (\ref{vfeconf}) of the vortex free energy 
implies confining behavior (`area-law') for the Wilson loop.

\subsection{Strong coupling cluster expansion and confinement}

We now return to our considerations at the end of section \ref{Z} 
regarding the flow of the coefficients  $\tilde{c}_j(n,
\alpha_{\Lambda,h}^{(n)}(t))$ in our partition function 
representations (\ref{A}) and (\ref{B}).  
This flow is bounded from above by that of 
the MK coefficients $c_j^U(m)$ regardless of the specific 
value assumed by the $\alpha_{\Lambda,h}^{(m)}(t_m)$'s at each decimation step 
(cf (\ref{cineq5})). Furthermore, by explicit evaluation under 
the iteration rules  (\ref{RG2}) - (\ref{RG5}), one finds that 
$c_j^U(n) \to 0$ 
as $n\to \infty$ for any initial $\beta$, provided $d\leq 4$.  
Thus, given any initial $\beta$, one may always take the number of iterations 
$n$ large enough  
so that the coefficients $c_j^U(n)$ become small 
enough to be within the region of convergence of the 
strong coupling expansion. Then by V.1: 
\beq 
\exp(-F_\Lambda^{(-)}(\beta)) = 
{ Z_{\Lambda^{(n)}}^{(-)}\,\Big(\{\,\tilde{c}_j(n,\alpha_\Lambda^{*\,(n)})\,\}
\Big) \over Z_{\Lambda^{(n)}}\,\Big(\{\, \tilde{c}_j(n,\alpha_\Lambda^{*\,(n)})
\,\}\Big) } \;. \label{vfeA} 
\eeq
The vortex free energy may then be evaluated in terms of the coefficients 
$\tilde{c}_j(n,\alpha_\Lambda^{*\,(n)})$ directly on lattice $\Lambda^{(n)}$ 
of spacing $b^n a$ within a convergent strong coupling polymer  
expansion.

Recall that, in the pure lattice gauge theory context, a polymer is 
a set $Y$ of connected plaquettes containing no `free' bond, i.e. no bond 
belonging to only one plaquette in $Y$ (see e.g. \cite{Mu}). 
The activity of a polymer $Y$ is defined by 
\beq 
z(Y)= \int\;\prod_{b\in Y} dU_b\;\prod_{p\in Y} g_p(U,n) \;, \label{zY}
\eeq
where 
\beq
g_p(U,n) = \sum_{j\not= 0} d_j\, \tilde{c}_j(n,\alpha_\Lambda^{*\,(n)})
\, \chi_j(U_p) \;.\label{g1}
\eeq
The polymer expansion is then 
\beq
\ln\dZ{n}  
= \sum_{X \subset \Lambda^{(n)}} \,a(X)\;\prod_{Y_i \in X} \;
z(Y_i)^{n_i} \;, \label{clusterexp1} 
\eeq
where the sum is over all linked clusters of polymers in $\Lambda^{(n)}$, 
each cluster $X$ consisting of a connected set of polymers 
$Y_i$, $i=1,\ldots,k_X$ with multiplicities $n_i$.  The 
combinatorial factor $a(X)$ is given by  
\beq
a(X)=\sum_{G(X)} (-1)^{l(G)} \, , \label{combfactor}
\eeq
where the sum is over all connected graphs on $X$ (full set (including 
multiplicities) 
$\{Y_i\}$ as vertices with a line connecting overlapping polymers) 
and $l(G)$ is the number of lines in the graph. 

In the case of $\dZ{n}^{(-)}$, the presence of the 
flux enters the activities 
(\ref{zY}) through the replacement (\ref{twist2}). We denote the resulting 
activities by $z^{(-)}(Y)$. This replacement 
does not affect polymers that are wholly contained in a simply 
connected part of $\Lambda^{(n)}$, since, in this case, the flux 
can be removed by a change of variables in the integrals in (\ref{zY}). 
Only clusters that contain at least one non-simply connected 
polymer forming a topologically non-trivially closed surface  
can be affected. Thus, one has 
\beq
\ln\dZ{n}^{(-)} - \ln \dZ{n}
= \sum_{X \subset \Lambda^{(n)}} \,a(X)\;\left(\,\prod_{Y_i \in X} \;
z^{(-)}(Y_i)^{n_i} - \,\prod_{Y_i \in X} \;
z(Y_i)^{n_i}\right)\;, \label{clusterexp2} 
\eeq
where the sum is only over all such topologically nontrivial  
linked clusters, the contribution of all other clusters 
canceling in the difference. The minimal cluster of this type consists 
of a single polymer which is a 
2-dimensional plane $\Pi: x_\mu=$const., $\mu=3,\ldots,d$ on $\Lambda^{(n)}$, 
thus of size $A^{(n)}=L_1^{(n)}L_2^{(n)}$, and activity  
\beq 
z(\Pi)=  \sum_{{\rm half-int.}\atop j\geq 1/2} 
\tilde{c}_j(n,\alpha_\Lambda^{*(n)})^{A^{(n)}} 
= \tilde{c}_{1/2}(n,\alpha_\Lambda^{*(n)})^{\,A^{(n)}} \,[\, 1+ 
\sum_{{\rm half-int.}\atop j\geq 3/2}
\left({\tilde{c}_j(n,\alpha^{*(n)}) \over \tilde{c}_{1/2}(n,\alpha^{*(n)})}
\right)^{A^{(n)}}\,] \;.\label{lead}
\eeq
(Note that the terms from the higher representations in (\ref{lead}) become 
utterly negligible in the large volume limit.) 
There are $L^{(n)}=\prod_{\kappa\not=1,2}L_\kappa^{(n)}$ such minimal 
clusters giving the leading contribution 
in (\ref{clusterexp2}). This leading contribution is thus 
seen to give the confining behavior (\ref{vfeconf}). 
Nonleading contributions come from nonminimal clusters  
consisting of $\Pi$ with or without `decorations', and  additional polymers 
touching $\Pi$. Such corrections have been evaluated 
in terms of the character expansion coefficients 
(the $\tilde{c}_j(n,\alpha_\Lambda^{*(n)})$'s in our case) to quite high order 
\cite{Mu}. They can be shown to exponentiate, so that 
\beq
{1\over L}\,F_\Lambda^{(-)}(\beta) = \exp(- \hat{\sigma}_\Lambda\,A) 
\eeq 
with 
\bea
\hat{\sigma}_\Lambda 
& = & {1\over b^{2n}}\,\kappa_\Lambda(n,\alpha_\Lambda^{*(n)}) \nonumber\\
& = & {1\over b^{2n}}\, \Big[\, \kappa(n,\alpha_\Lambda^{*(n)}) 
+ O\left( (\tilde{c}_{j+1/2}/\tilde{c}_{1/2})^{L_\mu^{(n)}}\right) 
+ O(n/A^{(n)}) \,\Big] \;, \label{sigma1}
\eea
where \cite{Mu}
\beq 
\kappa(n,\alpha_\Lambda^{*(n)}) =
\Big[-\ln \tilde{c}_{1/2}(n,\alpha_\Lambda^{*(n)})  
- 4\,\tilde{c}_{1/2}(n,\alpha_\Lambda^{*(n)})^4 +8\,\tilde{c}_{1/2}(n,
\alpha_\Lambda^{*(n)})^6 + \ldots \Big] \,. \label{sigma2}
\eeq
By the convergence of the expansion \cite{Ca}, the large volume limit  
exists and is given given by $\hat{\sigma}=\kappa(n,\alpha^{*(n)})/b^{2n}$, 
where $\alpha^{*(n)}$ is the lattice independent part of 
$\alpha_\Lambda^{*(n)}$ (cf. (\ref{alphsplit1})).

The number of iterations $n$ in the above expressions is taken large enough 
so that, given some initial $\beta$ on $\Lambda$, the resulting 
$c_j^U(n)$ are within the  
expansion convergence regime, and one can write the representation 
(\ref{vfeA}) by V.1. This implies the existence of a scale, a point to 
which we return below. Otherwise, $n$ is arbitrary. 
By construction, our procedure is such that the ratio (\ref{vfe}) 
is reproduced under successive decimations. 
Thus, given (\ref{vfeA}) at some $n$, suppose one performs one more 
decimation to lattice $\Lambda^{(n+1)}$. The condition that 
determines $\alpha_\Lambda^{*(n+1)}$ such that (\ref{vfeA}) is 
preserved is 
\beq
\kappa_\Lambda(n,\alpha_\Lambda^{*(n)}) = {1\over b^2}\,\kappa_\Lambda(n+1,
\alpha_\Lambda^{*(n+1)}) \;,\label{kappaI}
\eeq
which then results in constant string tension $\hat{\sigma}_\Lambda$ 
under successive decimations. 
Using (\ref{sigma2}) with (\ref{interc1}) and (\ref{RGstrong}),  
it is an easy exercise to solve (\ref{kappaI}), at least         
to leading approximation, for $\alpha^{*(n+1)}$. 
The $t$-parameter value $t_{h}^{(n+1)}$ this $\alpha^{*(n+1)}$ 
corresponds to can then also be easily obtained, if desired, from 
\[ h(\alpha^{*(n)},t)\ln F_0^U(n+1)\,|\Lambda^{(n+1)}| + 
\ln \dZ{n+1}\,\Big(\{\, \tilde{c}_j(n+1,\alpha^{*\,(n+1)})  
= \ln \dZ{n}\,\Big(\{\, \tilde{c}_j(n,\alpha^{*\,(n)}) 
\Big) \,,  \]
with $\ln\dZ{n}$, $\ln\dZ{n+1}$ given by (\ref{clusterexp1}) - in fact, to 
leading approximation, $\ln\dZ{n+1}$ can be ignored. 
Note that this amounts to replacing 
the set of the two equations (\ref{interI3A}) -
(\ref{interIfixplusA}) in V.1 by their ratio and 
one of them. This is indeed the most convenient procedure once 
(\ref{vfeA}) has been achieved.

\subsection{String tension and asymptotic freedom}

$\kappa(n,\alpha^{*(n)})$ is the string tension in lattice units of 
lattice $\Lambda^{(n)}$. It is a complicated,  
but well-defined function  of the original coupling $\beta= 4/g^2$ defined on 
lattice $\Lambda$,  eq. (\ref{Wilson}) (cf. remarks following 
(\ref{ratioder})). We write 
\beq
\kappa(n,\alpha^{*(n)}) \equiv \hat{\sigma}(n,g) \,.\label{sigma3}
\eeq
In dimensional units the asymptotic string tension in 
$d=4$ (\ref{sigma1}) - (\ref{sigma2}) is then 
\bea
\sigma  & = & {1\over a^2}
{1\over b^{2n}}\, \hat{\sigma}(n,g)
\label{sigma4a} \\
& = & {1\over a^2} \, \hat{\sigma}(g) \,. \label{sigma4b}
\eea
Here, as remarked above, $n$ is assumed greater than some required smallest  
$n(g)$.   
This (dynamically generated) physical scale, or some chosen multiple of 
it, is the only parameter in 
the theory. Fixing it specifies how the coupling $g$ must vary 
with changes of the (unphysical) lattice spacing $a$.

It is convenient, and customary, to introduce a fixed scale $\bL$ 
serving as an arbitrary unit of physical scales. Setting      
\beq
\bL^{\:-1} = a b^n \, ,\label{length}
\eeq 
determines the lattice spacing $a$ such that it takes $n$ steps to 
reach length scale  $1/\bL$:   
\beq 
n= {1\over \ln b} \ln {1\over a\, \bL} \,. \label{n-a}
\eeq
Fixing the string tension, given in units of $\bL$:  
\beq
\sigma = k \bL^2 \;,\label{sigma5}
\eeq
implies   
\beq 
\hat{\sigma}(n,g) = k \label{sigmaI} 
\eeq 
for some constant $k$. 
(\ref{sigmaI}) specifies the dependence of the bare coupling 
$g$ on $n$, hence, through (\ref{n-a}), the dependence on the 
lattice spacing $a$. It gives then the value $g(a)$ specified by the 
value of the string tension. (This is, of course, equivalent to fixing 
(\ref{sigma4b}) directly.)

Since 
\[ \hat{\sigma}(n+1,g+\Delta g)^{1/2} - \hat{\sigma}(n,g)^{1/2} =
b\, [\,\hat{\sigma}(n,g+\Delta g)^{1/2} - \hat{\sigma}(n,g)^{1/2}\,] + 
(b-1) \hat{\sigma}(n,g)^{1/2} \]
and  $\Delta a = -(b-1)a/b$ for $\Delta n=1$, one has from 
(\ref{sigmaI}): 
\beq
{\Delta \sqrt{\hat{\sigma}} \over \Delta g } (a {\Delta g\over \Delta a}) 
= \sqrt{\hat{\sigma}}  \label{sigmaIdiff}
\eeq 
If $(a{\Delta g /\Delta a})\equiv\beta(g)$, the `beta-function', is known, 
(\ref{sigmaIdiff}) can be integrated directly for $\hat{\sigma}(g)$.  
(This introduces a dimensional integration constant which can serve as 
the scale $\bL$). 
This is in fact the familiar 
textbook argument were one {\it assumes} the existence of a 
string tension so as to get (\ref{sigmaIdiff}), in which the standard weak 
coupling perturbative expression for the beta function is then used.

For us, however, the existence of a non-zero string tension is the 
outcome of the process of successive decimations to coarser scales as 
developed above. This process embodies all relevant information in the theory. 
In particular, it also supplies the specification of the function $g(a)$.

One can indeed construct the function $g(a)$ directly as follows:\\ 
(i) Starting with some initial value of $\beta=4/g^2$ perform 
successive decimations following the flow into the strong coupling 
regime with resulting string tension $\kappa(n,\alpha^{*(n))})$, 
eq. (\ref{sigma2}), at some $n=n_0$. Let $k$ denote the value of this 
string tension. The corresponding 
value of the lattice spacing $a_0$ is given by (\ref{n-a}), and $g=g(a_0)$. \\
(ii) Fix the string tension as in (\ref{sigmaI}). This is then satisfied at  
$n_0$, $g(a_0)$. \\
(iii) Vary $g$ away from $g(a_0)$ to determine $g$ such that, under  
successive decimations following the flow into the strong coupling 
regime, the resulting string tension satisfies (\ref{sigmaI}) for 
$n=n_0+1$.\\ 
(iv) Repeat (iii) for $n=n_0+2, n_0+3,\cdots, n_0-1, \cdots$. \\
This provides the functional relation $g(a)$. In particular, for 
$b=2$, it gives the sequence of values $g(a_0/2^l)$, $l=1,2,\ldots$,  
starting from some value $g(a_0)$.\footnote{This is the analog in the 
present context of the `staircase' procedure in  \cite{Cr}.}

Note that, according to (i) above,  
the number of decimations $n_0$ at which one chooses to apply 
V.1 to obtain (\ref{vfeA}), (\ref{sigma2}) amounts to fixing the string 
tension. This is the only physical parameter in the theory. 
A specification of $\bL$ is a specification of the value 
$g(a_0)$ at spacing $a_0$, which is a 
convention of no physical import.

One then has in principle a constructive method for obtaining 
$g(a)$ by a sequence of simple algebraic operations.
This is the coupling $g(a)$ as defined in the physical 
non-perturbative renormalization scheme 
specified by keeping the string tension fixed. 

A straightforward illustration of the method is provided by setting all 
$\alpha_\Lambda^{(n)}=1$, i.e. apply it to the flow according to the  
upper bound coefficients $c_j^U$ in (\ref{S1}). This yields   
$g(a)$ as given by MK decimations. 
We cannot apply it explicitly to the 
case of interest, i.e. the flow following the middle column 
coefficients in (\ref{S2}), 
since we do not determine them explicitly in this paper.   
The qualitative features at strong and weak coupling, however, are readily 
discernible.   

At strong coupling, i.e. small initial $\beta$, the number 
of decimations needed to reach a given string tension is of 
order unity, i.e. the lattice spacing $a$ is large: $a = O(\bL^{\,-1})$, 
and one is very far from any continuum limit.  
Successive decimations, by  construction, reproduce the behavior seen 
within the strong coupling expansion, and 
the familiar strong coupling variation given by $\beta(g) \sim g \ln g$ is 
the result, as can be checked by a short computation.

The opposite limit of large initial $\beta$ corresponds to large 
number of decimations, hence $a \ll \bL^{\,-1}$. Indeed, recall that 
$g=0$ is a fixed point of the decimations. Hence, for $\beta\to \infty$,  
one necessarily has $n\to \infty$ in order for, say, the leading upper bound 
coefficient $c_{1/2}^U(n)$ to reach any prescribed value $ < 1$. 
Thus, $a\bL \to 0$. Note that this limit is well-defined by construction 
since everything is bounded and continuous under successive 
decimations. 
Asymptotic freedom, i.e. the statement 
that $g(a)\to 0$ as $a\to 0$, is then a direct qualitative 
consequence of the flow produced by the decimations.  

It is instructive to examine the actual manner in which $g(a)\to 0$ 
under the upper bound decimations, i.e. the 
$c_j^U(m)$'s in (\ref{S2}). Comparing two $g$ values that differ by 
one decimation step ($b=2$), one finds 
\beq
{1\over g^2(a)} ={1\over g^2(2a)} + 2b_0 \,\ln2 + O(g^2) 
\eeq
for sufficiently small $g(a)$. The constant $b_0=(1-1/b^2)/(24\ln b)$ 
underestimates   
the value $11/24\pi^2$ obtained in a continuum perturbative calculation 
by only about $3\%$. 

The actual flow (middle column in (\ref{S2})) is faster, 
corresponding to somewhat larger $b_0$. According to RG lore, 
a beta-function defined by other means, such 
as fixing some renormalized coupling within weak coupling perturbation 
theory, should coincide, in its universal 
first two terms, with that defined by the above physical non-perturbative 
scheme. This, however, is outside the scope of, and 
not of direct relevance for the main argument in this paper. 

To reiterate, the above procedure 
completely specifies the dependence $g(a)$ in the physical 
renormalization scheme defined by keeping the 
string tension fixed, and this dependence is necessarily such 
that $g(a)\to 0$ as $a\to 0$.

\section{Concluding remarks}\label{SUM}

In summary, we obtained a representation of the vortex free energy, originally 
defined on a lattice of spacing $a$,   
in terms of partition functions on a lattice of spacing $ab^n$. 
The effective action in this representation 
is bounded by the corresponding effective action resulting 
from potential moving decimations (MK decimations) from spacing $a$ to 
spacing $ab^n$. The latter are explicitly computable. Confining behavior 
is the result, starting from any initial coupling $g$ on spacing $a$, 
by taking the number of decimation $n$ large enough. 

It is worth remarking again that in an 
approach based on RG decimations the fact that the only 
parameter in the theory is a physical scale emerges in a natural way. 
Picking a number of decimations can be related to 
fixing the string tension. That this can 
be done only after flowing into the strong coupling regime  
reflects the fact that this dynamically generated scale is an `IR effect'. 
The coupling $g(a)$ is completely determined in its dependence on $a$
once the string tension is fixed. 
In particular, $g(a) \to $ as $a\to 0$.
Note that this implies that there 
is no physically meaningful or unambiguous way of non-perturbatively viewing 
the short distance regime independently of the long distance regime.  
Computation of all physical observable quantities 
in the theory must then give a multiple of the string tension or 
a pure number. In the absence of other interactions, this scale provides 
the unit of length; there are in fact no free 
parameters.\footnote{This is part of the meaning of the common 
saying ``QCD is the perfect theory''.}

There is a variety of other results related to the approach in this 
paper that could not be included here. We note, in particular, 
that the same procedure  can be immediately transcribed to the 
Heisenberg $SU(2)$ spin model.  
Also, apart from analytical results, the considerations in this 
paper may be combined with Monte Carlo RG techniques to 
constrain the numerical construction of improved actions at different 
scales, a subject of perennial interest to the practicing lattice 
gauge theorist. We hope to report on these matters elsewhere.

This research was partially supported by 
NSF grant NSF-PHY-0555693.

\setcounter{equation}{0}
\appendix
\renewcommand{\theequation}{\mbox{\Alph{section}.\arabic{equation}}}
\section{Appendix}

In this appendix we obtain the lower and upper bounds 
II.1, IV.2, and III.1, III.2, IV.3, IV.4, and also IV.1.   

{\bf \S1.} To prove II.1 take the lattice $\Lambda^{(n)}$ to have length 
$L^{(n)}_\mu=2^{m_\mu}$, with 
integer $m_\mu$ in each direction $\mu=1,\cdots,d$, and torus topology 
(periodic boundary conditions) in all directions.  
Choose a hyperplane $\pi_1$\footnote{Actually a pair of hyperplanes because 
of the toroidal lattice topology. Following common 
practice, when a pair is actually meant, 
it will not be explicitly pointed out for brevity.}  
without sites 
perpendicular to, say, the $x^1$-direction and bisecting $\Lambda^{(n)}$, 
so that $\Lambda^{(n)}= \Lambda^{(n)}_L\cup \Lambda^{(n)}_R$ with 
$\Lambda^{(n)}_R = {\cal R}[\Lambda^{(n)}_L]$, where  ${\cal R}$ 
denotes reflection in $\pi_1$. Dropping now  the terms in (\ref{PF2}) 
coming from  all the plaquettes bisected by $\pi_1$, all non-negative by 
reflection positivity, gives 
\beq
Z_{\Lambda^{(n)}}(\{c_j(n)\}) = \int\, d\mu^0_{\Lambda^{(n)}}
\geq \left(\,\int \, d\mu^0_{\Lambda^{(n)}_L} \,\right)^2 \;.\label{ZL1}
\eeq
Note that the `half-lattice' $\Lambda^{(n)}_L$ has a boundary with resulting 
free boundary conditions for $d\mu^0_{\Lambda^{(n)}_L}$ in  the 
$x^1$-direction. ($d\mu^0_{\Lambda^{(n)}_L}$ still has periodic 
boundary conditions in all the other directions.) 
Next take a plane $\pi^\prime_1$ bisecting $\Lambda^{(n)}_L$ so that 
$\Lambda^{(n)}_L= \Lambda^{(n)\,\prime}_L\cup \Lambda^{(n)\,\prime}_R$ 
with 
$\Lambda^{(n)\,\prime}_R = {\cal R}^\prime[\Lambda^{(n)\,\prime}_L]$, where  
${\cal R}^\prime$ denotes reflection in $\pi_1^\prime$.
Removing now all plaquettes bisected by $\pi^\prime$ on the r.h.s. of 
(\ref{ZL1}) gives 
\beq
\int\, d\mu^0_{\Lambda^{(n)}}
\geq \left(\,\int \, d\mu^0_{\Lambda^{(n)\,\prime}_L} \,\right)^4 \, .
\label{ZL2}
\eeq
Proceeding in this manner one arrives at 
\beq 
\int\, d\mu^0_{\Lambda^{(n)}}
\geq \left(\,\int \, d\mu^0_{\Lambda^{(n)}_L(1)} \,\right)^{|L^{(n)}_1|}
 \;. \label{ZL3}
\eeq
In (\ref{ZL3}) $\Lambda^{(n)}_L(1)$ denotes  the lattice resulting from 
$\Lambda^{(n)}$ by reducing its extent to one lattice spacing in    
the $x^1$-direction, and $d\mu^0_{\Lambda^{(n)}_L(1)}$ is computed with 
free boundary conditions in the 
the $x^1$-direction. One next chooses a hyperplane normal to one of the 
remaining directions bisecting $\Lambda^{(n)}_L(1)$. Iterating this 
procedure in each successive direction, one eventually arrives at 
\beq 
\int\, d\mu^0_{\Lambda^{(n)}}
\geq \left(\,\int \, d\mu^0_h \,\right)^{|\Lambda_{(n)}|}
 \;,\label{ZL4}
\eeq
where $h$ denotes a hypercube, i.e. a subset of $L^{(n)}$ with $2^d$ 
sites. But, as it is easily seen, 
\beq 
\int \,d\mu^0_h  > \Big[\,1 + \sum_{j\not=0} d_j^2\,
c_j(n)^6 \,\Big]\,.\label{ZL5}
\eeq
Inserting (\ref{ZL5}) in (\ref{ZL4}) completes the proof of II.1. 
Note that, since the number of bonds is larger than the number of 
plaquettes, $|\Lambda^{(n)}|$ in (\ref{ZL4}) may be replaced by the latter.

\noindent{\bf \S2.} To obtain IV.2 let 
\beq
P^+_{\;\V}= {1\over 2}\,\left[\, 1 + \prod_{p\in \V}{f^{(-)}(U_p,n)\over 
f(U_p,n)}\,\right] = {1\over 2}\,\left[ 1 + \prod_{p\in \V} 
\exp \,\left[ \, A^{(-)}(U_p,n) - A(U_p,n)\,\right] \,\right] \,.\label{Pplus}
\eeq
Then 
\bea
Z^+_{\Lambda^{(n)}}(\{c_j(n)\}) & = & 
 \int \, d\mu^0_{\Lambda^{(n)}} \;
P^+_{\;\V}  \label{ZPplus1} \\
& = &  \int \, d\mu^0_{\Lambda^{(n)}} \;P^+_{\;\V}\; P^+_{\;\V^{\,\prime}}
\,, \label{ZPplus2} 
\eea 
where in the second equality $\V^{\,\prime}$ is any other 
coclosed plaquette set homologous to $\V$ (cf. remarks preceding eq. 
(\ref{twist2})). As in the text, we take $\V$ to wind around the lattice 
in the directions perpendicular to $x^1$- and $x^2$-directions. 
The measures defined by $\dZ{n}$ and $\dZ{n}^{(-)}$, and hence $\dZ{n}^+$,  
are clearly reflection positive in hyperplanes perpendicular to any one of 
the direction $\mu\not=1,2$. (\ref{ZPplus2}) makes it clear that the measure 
in $\dZ{n}^+$ is also reflection positive in planes normal to $\mu=1$ or 
$\mu=2$: simply take $\V^{\,\prime} = {\cal R} [\V]$ where ${\cal R}$ 
denotes reflection in such a plane. 
Thus, the measure in $\dZ{n}^+$ possesses RP in all planes. IV.2\,(i) is 
a trivial consequence of this fact. 

To establish IV.2\,(ii)
let a hyperplane $\pi$ normal to the $x^1$-axis bisect the lattice 
$\Lambda^{(n)}$
so that $\Lambda^{(n)}= \Lambda^{(n)}_L\cup \Lambda^{(n)}_R$. Take   
$\V \subset \Lambda^{(n)}_L$, and $\V^{\,\prime} = {\cal R} [\V] \subset 
\Lambda^{(n)}_R$. Dropping the terms in (\ref{ZPplus2}) coming from 
all plaquettes bisected by $\pi_1$, all of which are non-negative by RP, 
gives then
\beq
Z^+_{\Lambda^{(n)}}(\{c_j(n)\}) 
\geq \left(\,\int \, d\mu^0_{\Lambda^{(n)}_L} \;P^+_{\;\V}
\,\right)^2 \;.\label{Zplus1}
\eeq
But $\Lambda^{(n)}_L$ has a boundary with resulting 
free boundary conditions for $d\mu^0_{\Lambda^{(n)}_L}$ in  the 
$x^1$-direction. This implies that 
\beq
\int \, d\mu^0_{\Lambda^{(n)}_L} \;P^+_{\;\V} 
= \int \, d\mu^0_{\Lambda^{(n)}_L} \, ,
\eeq
since one may, by a shift of integration variables, move the 
location of the twist-carrying set $\V$ to this boundary and, by virtue 
of the free boundary conditions there, remove it from  $\Lambda^{(n)}_L$.  
The rest of the argument then proceeds exactly as in {\bf \S1} above. 
One thus arrives at (\ref{PFpluslowerb1}).

\noindent{\bf \S3.} The lower bounds III.2 and IV.4 are simple corollaries 
of II.1 and IV.2. Let $|\Lambda^{(n)}|$ denote the number of plaquettes in 
lattice $\Lambda^{(n)}$. One has 
\beq 
\Big[\,1 + \sum_{j\not=0} d_j^2\,
c_j(n-1)^6 \,\Big]^{|\Lambda^{(n-1)}|} 
\geq \Big[\,1 + \sum_{j\not=0} d_j^2\,
c_j(n-1)^6 \,\Big]^{|\Lambda^{(n)}|} \label{lessplaq}
\eeq
since $|\Lambda^{(n-1)}| > |\Lambda^{(n)}|$. So, by II.1(ii)
\bea 
\dZ{n-1}(\{c_j(n-1)\}) & \geq & 
\Big[\,1 + \sum_{j\not=0} d_j^2\,
c_j(n-1)^6 \,\Big]^{|\Lambda^{(n)}|}  \nonumber \\
 & \geq & \int dU_{\Lambda^{(n)}} \prod_{p\in \Lambda^{(n)}} \Big[\, 
1 + \sum_{j\not= 0} 
d_j\,c_j^L(n)\,\chi_j(U_p)\,\Big] \nonumber\\
&= &\dZ{n}( \{c_j^L(n)\} )\;, \qquad  \label{LA}
\eea
which gives III.2. 

Similarly, from (\ref{lessplaq}), IV.2(ii) and using (\ref{Z>Z-})
\bea 
\dZ{n-1}^+(\{c_j(n-1)\}) 
 & \geq & {1\over 2}\Bigg[\int dU_{\Lambda^{(n)}}  \prod_{p\in \Lambda^{(n})}
\Big[\, 1 + \sum_{j\not= 0} 
d_j\,c_j^L(n)\,\chi_j(U_p)\,\Big]  \nonumber \\
& & \qquad + 
\int dU_{\Lambda^{(n)}}  \prod_{p\in \Lambda^{(n})}
\Big[\, 1 + \sum_{j\not= 0} (-1)^{2jS_p[\V]}
d_j\,c_j^L(n)\,\chi_j(U_p)\,\Big] \Bigg]\nonumber \\
& = & \dZ{n}^+( \{c_j^L(n)\} )\;,    \label{LplusA}
\eea 
which gives IV.4. Note that, as it clear from (\ref{lessplaq}), both 
(\ref{LA}) and (\ref{LplusA}) are strict inequalities except in the 
trivial case where all $c_j(n-1)$ vanish. 

\noindent{\bf \S4.}  To obtain III.1 consider the decimation 
operation on the partition function 
$\dZ{m}(\{c_j(m)\})$ of the form (\ref{PF2}). Let $Q_{\mu\nu]}$ denote 
the set of all 
$[\mu\nu]$-plaquettes in $\Lambda^{(m)}$. Consider the set of all 
$3$-cells in $\Lambda^{(m)}$ 
of side length $b$ (in units of lattice spacing) in the $\kappa$-direction 
and unit side length in the $\mu$- and $\nu$-directions. 
The basic moving operation consists of moving the $(b-1)$ 
interior $[\mu\nu]$-plaquettes of each cell along the positive 
$\kappa$-direction to the location of the $[\mu\nu]$-plaquette 
in the boundary of the cell (Figure \ref{Dec1fig}). Then $Q_{[\mu\nu]}= 
Q_{[\mu\nu]}^- \cup Q_{[\mu\nu]}^+$, where $Q_{[\mu\nu]}^-$ is the 
set of all moved 
$[\mu\nu]$-plaquettes and $Q_{[\mu\nu]}^+$ the set of all `receiving' 
$[\mu\nu]$-plaquettes on the $3$-cell boundaries. The action of the receiving 
boundary plaquettes is renormalized as in (\ref{potmove2}). 
Given $\dZ{m}(\{c_j(m)\})$ with corresponding action $A_p(U,m)$, eq. 
(\ref{actdef}), define the action  
\beq 
A_{\Lambda^{(m)}}(U,m, \xi) = \sum_{p\in \Lambda^{(m)}} 
\;  A_p(U_p,m) + 
\sum_{p\in Q_{[\mu\nu]}^+}\;\xi\,(\zeta_0-1)\,A_p(U_p,m) - 
\sum_{p\in Q_{[\mu\nu]}^-}\;\xi\,A_p(U_p,m) \label{xiact}
\;,
\eeq
interpolating between the action before ($\xi=0$) and after ($\xi=1$) 
the move. Then 
\bea
{d\over d\xi} \dZ{m}(m, \xi) |_{\xi=0}  & \equiv & 
{d\over d\xi}\int dU_{\Lambda^{(m)}} \exp A_{\Lambda^{(m)}}(U,m,\xi)\;_{|\xi
=0} \label{firstder}\\
& = & 
(\zeta_0-1)\sum_{p\in Q_{[\mu\nu]}^+}\vev{A_p(U_p,m)}_{0,\,\Lambda^{(m)},
\,\xi=0}
-\sum_{p\in Q_{[\mu\nu]}^-}\vev{A_p(U_p,m)}_{0,\,\Lambda^{(m)},\,\xi=0}, 
\nonumber 
\eea  
where $<->_{0,\,\Lambda^{(m)},\,\xi}$ denotes the unweighted expectation 
with measure defined by (\ref{xiact}).  
Now, given a $p\in {\cal Q}_{[\mu\nu]}^{\pm}$, one notes 
that $\vev{A_p(U_p,m)}_{0,\,\Lambda^{(m)},\,\xi=0}$ in (\ref{firstder}) 
does not vary with $p$ along the $\kappa$-direction 
i.e. along the direction of the move. Hence, if 
\beq 
\zeta_0= b \label{MKchoice}
\eeq
(\ref{firstder}) gives 
\beq
{d\over d\xi} Z_\Lambda(m, \xi) |_{\xi=0} = 0 \;, \label{firstder1} 
\eeq
since there are $(b-1)$ plaquettes in $Q_{[\mu\nu]}^-$ 
for each plaquette in $Q_{[\mu\nu]}^+$.  
But   
\beq
{d^2\over d\xi^2} \dZ{m}(m, \xi) 
= \vev{\left( \sum_{p\in Q_{[\mu\nu]}^+}\;(\zeta_0-1)\,A_p(U_p,m) - 
\sum_{p\in Q_{[\mu\nu]}^-}\;A_p(U_p,m) \right)^2}_{0,\,\Lambda^{(m)},\xi}  
\geq 0 \;.  \label{secondder}
\eeq 
(\ref{firstder1})-(\ref{secondder}) then imply that, with condition 
(\ref{MKchoice}), 
$\dZ{m}(m, \xi)$, and\footnote{Indeed,  
the second derivative of $\ln \dZ{m}(m,\xi)$ w.r.t. $\xi$ is 
also positive by the usual convexity of the free energy, a statement 
that follows generally by an application of H\"{o}lder's inequality.} 
$\ln \dZ{m}(m,\xi)$, are increasing convex 
functions of $\xi$ on $0\leq \xi\leq 1$. Thus, $\dZ{m}(m,0)\leq \dZ{m}(m,1)$. 

\begin{figure}[ht]
 \begin{center}\resizebox{6cm}{!}{\input{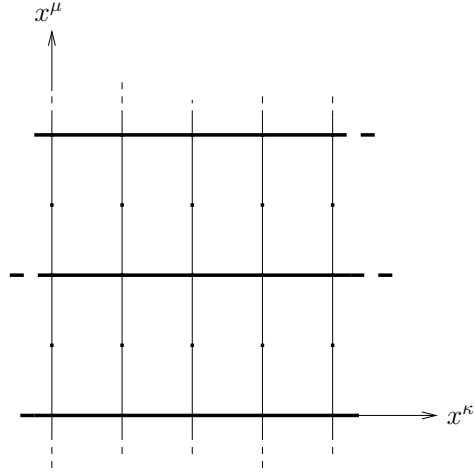}}
\end{center}
 \caption{\label{ADec2fig}Translational invariance along undecimated direction 
$x^\kappa$ at $\xi=0$ (see text) after decimation along $x^\mu$.
 }
\end{figure}

A complete decimation $\Lambda^{(m)} \to \Lambda^{(m+1)}$ is 
performed by repeating the above basic moving-renormalization step 
in each of the available normal directions for each possible plaquette 
orientation as described in section \ref{DEC1}. 
One need only observe that, in carrying out each such successive 
step, translation invariance at $\xi=0$ for a given plaquette along 
the direction of the move holds regardless of any previous moves 
performed along other directions as schematically depicted in Figure 
\ref{ADec2fig}. 
Each step then with $\zeta_0=b$ results in a further upper bound on $\dZ{m}$. 
The completion of all the moving-renormalization steps results into 
(\ref{U}) at $r=1$, i.e. 
\beq
\dZ{m}(\{c_j(m)\})\,  \leq \,  
F_0^U(m+1)^{|\Lambda^{(m+1)}|}\, \dZ{m+1}(\{c_j^U(m+1,1)\})\;,\label{UA}
\eeq 
with $F_0^U(m+1)$, $c_j^U(m+1,1)$ given by (\ref{RG2}) - (\ref{RG5}) 
at $\zeta=\zeta_0^{(d-2)}=b^{(d-2)}$. Furthermore, it follows 
from II.1(i), (\ref{upperc1}) and the fact that $0\leq c_j^U(m+1,1)\leq 1$  
that replacing $c_j^U(m+1,1)$ by $c_j^U(m+1,r)$ in the r.h.s. in (\ref{UA}) 
gives a decreasing function in $r$ on $0< r\leq 1$. 
This completes the proof of III.1.

The proof of III.1 can be used essentially unaltered to obtain IV.3. 
Introducing the analog 
of (\ref{xiact}) for the action (\ref{PF1btwist}) in the presence of flux:  
\beq 
A_{\Lambda^{(m)}}^{(-)}(U,m, \xi) = \sum_{p\in \Lambda^{(m)}} 
\;  A_p^{(-)}(U_p,m) + 
\sum_{p\in Q_{[\mu\nu]}^+}\;\xi\,(\zeta_0-1)\,A_p^{(-)}(U_p,m) - 
\sum_{p\in Q_{[\mu\nu]}^-}\;\xi\,A_p^{(-)}(U_p,m) \label{xiacttwist}
\;,
\eeq
one has 
\beq
{d\over d\xi} \dZ{m}^{(-)}(m, \xi) |_{\xi=0}  = 
(\zeta_0-1)\sum_{p\in Q_{[\mu\nu]}^+}\,
\vev{A_p^{(-)}(U_p,m)}_{0,\,\Lambda^{(m)},\,\xi=0}^{(-)}
-\sum_{p\in Q_{[\mu\nu]}^-}\;
\vev{A_p^{(-)}(U_p,m)}_{0,\,\Lambda^{(m)},\,\xi=0}^{(-)}\; , 
\label{firstdertwist}
\eeq
where $<->_{0,\,\Lambda^{(m)},\,\xi}^{(-)}$ denotes the unweighted expectation 
with action (\ref{xiacttwist}).   
The observation that, having moved plaquettes in certain directions,  
there is still translational 
invariance at $\xi=0$ for plaquettes to be moved in the remaining 
undecimated directions (Figure \ref{ADec2fig}) 
holds also in the presence of the flux. This is a consequence of the 
basic property (cf. section \ref{TZ}) that $\dZ{m}^{(-)}$ does not depend 
on the location but only the homology class of $\V$. One may, for example, 
always bring the set $\V$ by a change of variables to occupy  
the exact same location with respect to the plaquette $p$ in 
each expectation $\vev{A_p^{(-)}(U_p,m)}_{0,\,\Lambda^{(m)},\,\xi=0}^{(-)}$ in 
(\ref{firstdertwist}).   Thus it is again the case that each such  
expectation in (\ref{firstdertwist}) does not vary with 
$p\in Q_{[\mu\nu]}^{\pm}$ along the undecimated directions normal to $p$. 
Hence, when  (\ref{MKchoice}) holds, 
\[
{d\over d\xi} \dZ{m}^{(-)}(m, \xi) |_{\xi=0}=0 \;, 
\qquad \mbox{whereas} \qquad {d^2\over d\xi^2} \dZ{m}^{(-)}(m, \xi) 
\geq 0 \;,  \]
which gives the analog of (\ref{UA}) in the presence of flux:  
\beq
\dZ{m}^{(-)}(\{c_j(m)\}) \,  \leq \,  
F_0^U(m+1)^{|\Lambda^{(m+1)}|}\, \dZ{m+1}^{(-)}(\{c_j^U(m+1,1)\})\;.
\label{UAtwist}
\eeq 
Combining (\ref{UAtwist}) with (\ref{UA}) and IV.2(i), (\ref{upperc1}) 
then gives IV.3.

\noindent{\bf \S5.} To establish IV.1, take $\V$ such that all $p\in \V$ are 
bisected by a hyperplane $\pi$ perpendicular to, say, $x^1$. Let 
\beq 
S_p^{\ssc 1/2} = \sum_{j=\rm half-int.} d_j\,c_j(n)\,\chi_j(U_p) \;\, ,
\qquad \quad S_p^{\ssc1} = 1 + 
\sum_{j=\rm int.\atop j\not=0} d_j\,c_j(n)\,\chi_j(U_p) 
\eeq
denote the sums over half-integer and integer representations, 
respectively. Now 
\beq 
  \prod_{p\in \V} \Big[\, S_p^{\ssc1/2} + S_p^{\ssc 1} \,\Big] 
- \prod_{p\in \V} \Big[\, - S_p^{\ssc 1/2} + S_p^{\ssc 1} \,\Big] 
= 2 \!
\sum_{Q\subset \V \atop |Q|=\rm odd} \prod_{p\in Q} S_p^{\ssc 1/2} 
\prod_{p\in\V\setminus Q} S_p^{\ssc 1} \,,\label{Vexpand}
\eeq
where the sum is over all subsets of plaquettes $Q$ in $\V$ with 
odd number of plaquettes $|Q|\geq 1$. Inserting (\ref{Vexpand}) in 
$(\dZ{n} - \dZ{n}^{(-)})$ one has  
\beq
\dZ{n} - \dZ{n}^{(-)} = 2\! 
\sum_{Q\subset \V \atop |Q|=\rm odd} 
\int dU_{\Lambda^{(n)}} \prod_{p\in \Lambda^{(n)}\setminus \V} 
\Big[\, 1 + \sum_{j\not= 0} d_j\,c_j^L(n)\,\chi_j(U_p)\,\Big] 
 \prod_{p\in Q} S_p^{\ssc 1/2} \prod_{p\in
\V\setminus Q} S_p^{\ssc 1}  \,. \label{VexpandZZ-} 
\eeq
Since $c_j(n)\geq 0$, all $j$, every term in the sum in (\ref{VexpandZZ-}) 
is manifestly non-negative by RP in $\pi$, which proves IV.1.

\section{Appendix} 
\setcounter{equation}{0}

In this Appendix we give some simple estimates concerning the variation 
of the level surfaces $\alpha_{\Lambda,h}^{(m)}(t,r)$ and 
$\alpha_{\Lambda,h}^{+(m)}(t,r)$ w.r.t. the parameters 
$t,r$.

\noindent{\bf \S1.}  The derivative of 
$\alpha_{\Lambda,h}^{(m)}(t,r)$ 
w.r.t. $r$ is given by 
\beq
{\partial  \alpha_{\Lambda,h}^{(m)}(t,r)\over \partial r} = - \left[
{ \D  B_{\Lambda^{(m)}}(\alpha, r)
\over{\D  {\partial h(\alpha,t)\over \partial \alpha} + 
A_{\Lambda^{(m)}}(\alpha, r)} }\right]_{\alpha_{\Lambda,h}^{(m)}(t,r)} , 
\label{alphrder1}
\eeq
where 
\beq
B_{\Lambda^{(m)}}(\alpha, r) \equiv  {1\over \ln F_0^{U}(m) }\,
{1\over |\Lambda^{(m)}| }\, 
{\partial \over \partial r }\ln Z_{\Lambda^{(m)}}\,\Big(\{\tilde{c}_j( m, 
\alpha, r)\}\Big) <  0\;. \label{alphrder2}
\eeq

The derivative of $\alpha_{\Lambda,h}^{+(m)}(t,r,r^\prime)$ w.r.t. $r$ is 
similarly given by 
\beq
{\partial  \alpha_{\Lambda,\,h^+}^{+(m)}(t,r)\over \partial r} = 
- \left[{ \D  B^+_{\Lambda^{(m)}}(\alpha, r)
\over{\D  {\partial h^+(\alpha,t)\over \partial \alpha} + 
A^+_{\Lambda^{(m)}}(\alpha, r)} }\right]_{\alpha_{\Lambda,\,
h^+}^{+(m)}(t,r)} \label{alphplusrder1}
\eeq  
with
\beq
B^+_{\Lambda^{(m)}}(\alpha, r) \equiv  {1\over \ln F_0^{U}(m) }\,
{1\over |\Lambda^{(m)}| }\, 
{\partial \over \partial r }\ln Z^+_{\Lambda^{(m)}}\,\Big(\{\tilde{c}_j( m, 
\alpha, r)\}\Big) <  0\;. \label{alphplusrder2}
\eeq

\noindent{\bf \S2.} In this Appendix we use the short-hand notation
$Z_{\Lambda^{(m)}}=  Z_{\Lambda^{(n)}}(\{\tilde{c}_j(m,\alpha,r)\})$ 
and $\dZ{m}^+= \dZ{m}^+(\{\tilde{c}_j(m,\alpha,r)\})$; and   
also, given any set of plaquettes $P \subset \Lambda^{(m)}$, we define:    
\beq 
d\tilde{\mu}^0_{P} \equiv \prod_{b\in \Lambda^{(m)}}\,dU_b\;
\prod_{p\in P}f_p(U_p,m,\alpha,r)  \label{shortdmu} \, .
\eeq

By translational invariance 
\beq
A_{\Lambda^{(m)}}(\alpha, r) = 
{1\over \ln F^U_0(m)}\,\sum_{j\not=0} d_j \,{\partial \tilde{c}_j(m,\alpha, 
r)\over \partial \alpha} \,{1\over Z_{\Lambda^{(m)}}} 
\int d\tilde{\mu}^0_{\Lambda^{(m)}\setminus p}\; \chi_j(U_p) \,. \label{Aest1} 
\eeq
Since 
\[ Z_{\Lambda^{(m)}} \geq  
\int d\tilde{\mu}^0_{\Lambda^{(m)}\setminus p}\; \chi_j(U_p) \geq 0 \] 
by RP in a plane  bisecting $p$, one has 
\bea
A_{\Lambda^{(m)}}(\alpha, r) & \leq & 
 {1\over \ln F^U_0(m)}\,\sum_{j\not=0} 
d_j^2 \; {\partial \tilde{c}_j(m,\alpha, 
r,r^\prime)\over \partial \alpha} \nonumber \\
  & = & {1\over \ln F^U_0(m)}\,||{\partial g_p(m,\alpha,r)\over 
\partial \alpha}|| \,. \label{Aest2}
\eea 
In the same manner one obtains 
\bea 
|B_{\Lambda^{(m)}}(\alpha, r)|  & \leq & 
{1\over \ln F^U_0(m)}\,\sum_{j\not=0} 
d_j^2 \; |{\partial \tilde{c}_j(m,\alpha, 
r)\over \partial r}| \nonumber \\
& = & {1\over \ln F^U_0(m)}\,||{\partial g_p(m,\alpha,r)\over 
\partial r}|| \,. \label{Best1} 
\eea

By translational invariance for plaquettes of the same orientation in 
the presence of flux, the same bounds can similarly be shown to hold for 
the quantities 
$A^+_{\Lambda^{(m)}}$ and $B^+_{\Lambda^{(m)}}$: 
\beq   
 A^+_{\Lambda^{(m)}}(\alpha,r) \leq 
{1\over \ln F^U_0(m)}\,||{\partial g_p(m,\alpha,r)\over 
\partial \alpha}||  \label{Aest2plus}
\eeq 
and 
\beq
|B^+_{\Lambda^{(m)}}(\alpha,r)| \leq 
{1\over \ln F^U_0(m)}\,||{\partial g_p(m,\alpha,r) \over 
\partial r}||  \,. \label{Best1plus} 
\eeq
Note that all these upper bounds are independent of the lattice 
size $|\Lambda^{(m)}|$.

Let $p$ be a fixed plaquette, and $c$ a 3-cube having $p$ in its boundary 
$\partial c$ and 
protruding from $p$ in one of the $d-2$ directions normal to $p$. Let 
$Q$ denote the set of plaquettes sharing a bond with $\partial c$ 
but not belonging to $\partial c$. Then 
\bea 
{1\over Z_{\Lambda^{(m)}}} 
\int d\tilde{\mu}^0_{\Lambda^{(m)}\setminus p}\; d_j\,\chi_j(U_p) & = & 
{1\over Z_{\Lambda^{(m)}}}  
\int d\tilde{\mu}^0_{\Lambda^{(m)}\setminus Q\cup \partial c}\; 
d_j\,\chi_j(U_p)
\,\prod_{p^\prime \in Q\cup \partial c\setminus p} f_{p^\prime}(U_{p^\prime},
m,\alpha,r) \nonumber \\
& \geq & {1\over Z_{\Lambda^{(m)}}} 
\int d\tilde{\mu}^0_{\Lambda^{(m)}\setminus Q\cup \partial c}
\; d_j\,\chi_j(U_p) 
\, \prod_{p^\prime \in \partial c\setminus p} f_{p^\prime}(U_{p^\prime},m,
\alpha,r)\nonumber \\
& = & {1\over Z_{\Lambda^{(m)}}} 
\int d\tilde{\mu}^0_{\Lambda^{(m)}\setminus Q\cup \partial c}\;
d_j^2 \tilde{c}_j(m,\alpha,r)^5 \nonumber \\
& \geq & {1\over ||f_p(m,\alpha,r)||^{|Q|} } 
{d_j^2 \;\tilde{c}_j(m,\alpha,r)^5 \over 
[\,1+ \sum_{i\not=0} \, d_i^2 \;\tilde{c}_i(m,\alpha,r)^6 \,]} 
\,.\label{lowerAest1}
\eea
RP in each of the two planes  
bisecting the plaquette $p$ shows that the terms involving plaquettes 
in the set $Q$ in the first line are all positive which results in the 
inequality in the second line.  
Using (\ref{lowerAest1}) in (\ref{Aest1}) gives the lower bound 
\bea 
A_{\Lambda^{(m)}}(\alpha,r) &\geq & 
{1\over {||f_p||^{|Q|}\;\ln F^U_0(m)}}\,
\sum_{j\not=0} 
\,{\partial \tilde{c}_j(\alpha,r)\over \partial \alpha} 
{d_j^2 \tilde{c}_j(m,\alpha, r)^5
\over [\,1+ \sum_{i\not=0} \,d_i^2 \;\tilde{c}_i(m,\alpha,r)^6 \,] } 
\nonumber \\
&=&  {1\over ||f_p||^{|Q|} \; \ln F^U_0(m)}\;
\frac{1}{6}{\partial \over \partial \alpha} \,
\ln\left[\,1+ \sum_{j\not=0} \,d_j^2 \;\tilde{c}_j(m,\alpha,r)^6 \,
\right]  \,.\label{lowerAest2}
\eea
Similarly, one obtains a lower bound on  $|B_{\Lambda^{(m)}}|$ 
\bea 
|B_{\Lambda^{(m)}}(\alpha,r)| &\geq &  
{1\over ||f_p||^{|Q|} \; \ln F^U_0(m)}\,
\sum_{j\not=0} 
\,|{\partial \tilde{c}_j(\alpha,r)\over \partial r}| 
{d_j^2 \tilde{c}_j(m,\alpha, r)^5
\over [\,1+ \sum_{i\not=0} \,d_i^2 \;\tilde{c}_i(m,\alpha,r)^6 \,] } 
\nonumber \\
&=&  {1\over ||f_p||^{|Q|} \; \ln F^U_0(m)}\,
\frac{1}{6}(- {\partial \over \partial r}) \,
\ln\left[\,1+ \sum_{j\not=0} \, d_j^2 \;\tilde{c}_j(m,\alpha,r)^6 \,
\right] \label{lowerBest1} \,. 
\eea 
Again, these lower bounds are manifestly lattice-size independent.

Since, as it is easily seen (apply (\ref{ZPplus2})), (\ref{lowerAest1}) 
holds also when 
$d\mu^0_{\Lambda^{(m)}}$ is replaced by $d\mu^{0+}_{\Lambda^{(m)}}$, 
the r.h.s. of (\ref{lowerAest2}) and  
(\ref{lowerBest1}) also give lower bounds for 
$A^+_{\Lambda^{(m)}}$ and $|B^+_{\Lambda^{(m)}}|$, respectively.

\noindent{\bf \S3.} The lower bound in (\ref{collar}), i.e. 
\beq 
\delta^\prime < \alpha_{\Lambda,h}^{(m)}(t,r)
\label{collarBlow} 
\eeq
is a consequence of 
II.1 and (\ref{interI3}). From the first inequality in (\ref{LA}) 
(excluding the trivial case of all $c_j$ vanishing) one has 
\beq 
Z_{\Lambda^{(m-1)}}
>  \Big[\,1 + \sum_{j\not=0} d_j^2\,
c_j(m-1)^6 \,\Big]^{|\Lambda^{(m)}|} \,;\label{PFlowerb1B}
\eeq 
whereas from (\ref{interI3}), with the short-hand notation 
$\alpha_{\Lambda,h}^{(m)}(t,r)) = \alpha_{\Lambda,h}^{(m)}$,  
\beq
Z_{\Lambda^{(m-1)}} 
 \leq 
F_0^U(m)^{h_t(\alpha_{\Lambda,h}^{(m)})\,|\Lambda^{(m)}|}
\Big[ 1 + \sum_{j\not= 0} d_j^2\, 
\tilde{c}_j(m,\alpha_{\Lambda,h}^{(m)}, r)\Big]^{|\Lambda^{(m)}|} \;.  
\label{interI3upB}
\eeq
Combining (\ref{PFlowerb1B}) and (\ref{interI3upB}) gives 
\beq 
\Big[\,1 + \sum_{j\not=0} d_j^2\,
c_j(m-1)^6 \,\Big] < 
F_0^U(m)^{h_t(\alpha_{\Lambda,h}^{(m)})}
\Big[ 1 + \sum_{j\not= 0} d_j^2\, 
\tilde{c}_j(m,\alpha_{\Lambda,h}^{(m)}, r)\Big] \, ,\label{alphlowerb1}
\eeq
which, together with (\ref{hlimits}) and (\ref{lowerc2}) or 
(\ref{lowerc3}), shows that it cannot be that 
$\alpha_{\Lambda,h}^{(m)}(t,r) \to 0$ in any fashion with 
increasing lattice size $|\Lambda^{(m)}|$. 

An explicit lower bound on 
$\alpha_{\Lambda,h}^{(m)}$ is easily obtained 
from (\ref{alphlowerb1}) by taking (\ref{interc1}) with, for example,  
(\ref{lowerc3}) and $h_t(\alpha)= 
\alpha^{s(t)}, \; s(t)\geq 1$. 
Using the elementary inequality 
\beq
x^q - 1 \leq q (x-1)\;, \qquad x\geq 0\,, \quad 0\leq q \leq 1\;,
\label{elemineq}
\eeq 
(\ref{alphlowerb1}) gives 
\beq 
1 \; \geq \; \alpha_{\Lambda,h}^{(m)}(t,r) \; > \; 
{ \sum_{j\not=0} d_j^2\,c_j(m-1)^6  \over 
\left[ (\,F_0^U(m) -1\,) + F_0^U(m)\, \sum_{j\not= 0} d_j^2\, c^U_j(m,r)
\right] }  \equiv \delta^\prime \,> \, 0 \,.\label{alphlowerb2}
\eeq
Note that, from  (\ref{F0lower1D}) below,  
$(\,F_0^U(m) -1\,) > \sum_{j\not=0} d_j^2\,c_j^U(m-1)^2> 
\sum_{j\not=0} d_j^2\,c_j(m-1)^2$. Similar expressions can be obtained for 
other choices of $h_t$. 

The lower bound in (\ref{collarplus}) is similarly seen to hold by 
combining IV.2 and (\ref{interI3plus}).

To satisfy the upper bound  requirement in (\ref{collar}), i.e. 
\beq 
\alpha_{\Lambda,h}^{(m)}(t,r) < 1-\delta \;, \label{collarBup} 
\eeq 
it suffices to let the decimation parameter $r$ 
vary, if necessary, away from unity in the domain (\ref{rdomain}). 
To see this, suppose that, performing the m-th decimation step, one finds 
that $\alpha_{\Lambda,h}^{(m)}(t_m,1)= 1- \delta_\Lambda^{(m)}$, where, say, 
$\delta_\Lambda^{(m)}\leq O(1/|\Lambda^{(m)})$. 
Using the bounds (\ref{Aest2}), (\ref{lowerBest1}) 
on $A_{\Lambda^{(m)}}$, $B_{\Lambda^{(m)}}$, and  
the boundedness of the derivatives of 
$h(\alpha,t)$, in eq. (\ref{alphrder1}) then gives  
\beq
\partial \alpha_{\Lambda,h}^{(m)}(t,r)/\partial r \geq \theta >0 
\eeq 
for some constant $\theta$ independent of $|\Lambda^{(m)}|$ for all 
$0 < r\leq 1$.  
Hence, 
for $t\geq t_m$, and some $\xi$ between $0$ and $1$,   
\bea 
\alpha_{\Lambda,h}^{(m)}(t_m,1-\epsilon)& \leq & 
\alpha_{\Lambda,h}^{(m)}(t,1-\epsilon) \nonumber \\
& = &    \alpha_{\Lambda,h}^{(m)}(t,1) - \epsilon\;
{\partial \alpha_{\Lambda,h}^{(m)}(t,r) \over 
\partial r}\Bigg|_{r=1-\xi\epsilon} 
\nonumber \\
& \leq &  \alpha_{\Lambda,h}^{(m)}(t,1) - \epsilon\;\theta  \nonumber\\
& < & 1- \delta(\epsilon) 
\label{collarsize}
\eea
with $\delta(\epsilon) = \epsilon\theta/2$, and  (\ref{collarBup}) 
is satisfied.

Given (\ref{collar}), the bounds (\ref{Aest2}), (\ref{lowerAest2}) and the 
properties of the interpolation $h$ (cf. (\ref{hlimits}), (\ref{h})),  
it follows from (\ref{alphtder1}) and (\ref{htder}) that 
\beq 
{\D \partial \alpha_{\Lambda,h}^{(m)}\over \partial t} (t,r)
\geq  \eta_1(\delta) > 0 \, , \qquad \qquad 
- {\D d h\over \D dt}(\alpha_{\Lambda,h}^{(m)}(t,r),t)\geq 
\eta_2(\delta)>0  \, , \label{dercollarB} 
\eeq 
where $\eta_1$, $\eta_2$ are lattice-size independent, and 
$r$ in the domain (\ref{rdomain}).

Since the same upper and lower bounds apply to
$A^+_{\Lambda^{(m)}}$, $|B^+_{\Lambda^{(m)}}|$, 
the same considerations again 
show that eqs. (\ref{collarplus}) and 
(\ref{dercollarplus}) are always ensured to hold by letting 
$r$ vary, if necessary, in (\ref{rdomain}).

\section{Appendix} 
\setcounter{equation}{0}

\noindent{\bf \S1.} Under the conditions in V.1, given 
$\alpha_{\Lambda, h}^{(n)}(t)$ satisfying (\ref{interI3}),   
one seeks a solution $t=t^{(n)}_{\Lambda,h}$ to (\ref{interIfixplusA}), i.e. 
\beq  
\tilde{Z}^+_{\Lambda^{(n)}}\,(\beta, h, \alpha_{\Lambda, h}^{(n)}(t),\,t) 
=  Z_{\Lambda^{(n-1)}}^+\;.\label{interIfixplusC}
\eeq 
First note that, if a solution exists, it is unique since, as it is easily 
checked, $\tilde{Z}^+_{\Lambda^{(n)}}\,(\beta, h, 
\alpha_{\Lambda, h}^{(n)}(t),\,t)$ is monotonic in $t$.

To show that a solution exists, we proceed as follows. 
By (\ref{interI3plus}), given any suitable interpolation $h$, there is a 
function $\alpha_{\Lambda,\, h}^{+(m)}(t)$ such that 
\beq
\tilde{Z}^+_{\Lambda^{(n)}}\,(\beta, h, \alpha_{\Lambda,\, h}^{+(n)}(t),
\,t) =  Z_{\Lambda^{(n-1)}}^+  \label{interI3plusC}
\eeq
for all allowed values of the parameter $t$. We simplify notation 
in the following by omitting the 
fixed labels $n$, $\Lambda$, and write 
\[ \alpha_h(t) \equiv \alpha_{\Lambda, h}^{(n)}(t), \qquad 
\alpha_h^+(t)\equiv  \alpha_{\Lambda, h}^{+(n)}(t) \; . \]

\begin{figure}[ht]\begin{center}
\resizebox{11cm}{!}{\input{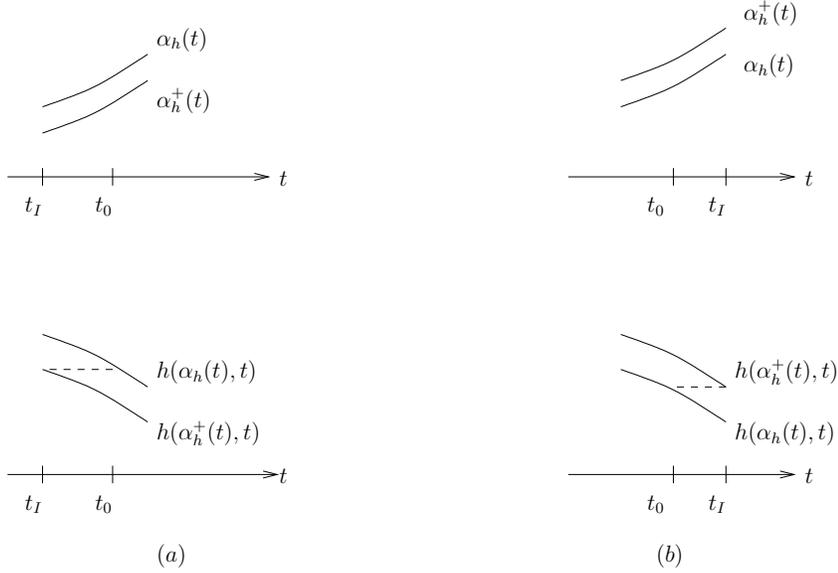}}
\caption{\label{C1abfig}Fixing $t_0$ in the neighborhood of some initial 
$t_I$. Distance between curves in the figure is greatly exagerated.}
\end{center} 
\end{figure}

Assume that there is a $t=t_I$ at which 
$\alpha_h(t_I)> \alpha_{h}^+(t_I)$. Then let 
$t_0$ be such that 
\beq
h(\alpha_h^+(t_I),\, t_I) =    
h(\alpha_h(t_0),\, t_0 ) \; , \qquad t_0\in [t_a,t_b] \;.
\label{htzero} 
\eeq
It is always possible to have such a $t_0$ by virtue of (\ref{dercollar}) 
and (\ref{dercollarplus}), which imply (Figure \ref{C1abfig}(a)) 
that $t_0 > t_I$ and 
\beq  
\alpha_h(t)> \alpha_h^+(t_0) > \alpha_h^+(t_I)\;,  \qquad 
 \qquad t\geq t_0\,. \label{alphtzero1}
\eeq 

With the notation 
\beq
\Phi_{\Lambda^{(n)}}^+(\alpha) \equiv 
{1\over \ln F_0^U(n)}{1\over |\Lambda^{(n)}|}
\ln \dZ{n}^+\big(\{\tilde{c}_j(n,\alpha)\}\big)  
\eeq
we now define 
\beq 
\Psi(\lambda, t)  \equiv 
h(\alpha_h(t), t) + (1-\lambda)\,\dPhi{n}^+(\alpha_h^+(t_I)) + 
\lambda\,\dPhi{n}^+(\alpha_h(t)) - \Phi^+_{\Lambda^{(n-1)}} \label{IM1} \;, 
\eeq
and consider the equation 
\beq 
\Psi(\lambda, t) =0 \;. \label{IM2}
\eeq
At $\lambda=0$ eq. (\ref{IM2}) is solved by setting $t=t_0$ since there, 
by (\ref{htzero}), it 
reduces to (\ref{interI3plusC}) evaluated at $t_I$. 
At $\lambda=1$, (\ref{IM2}) becomes the equation to be solved 
(\ref{interIfixplusC}).
A solution to (\ref{IM2}) then determines implicitly a function $t(\lambda)$ 
with the property $t(0)=t_0$. If this function can be 
extended on $0\leq \lambda \leq 1$, it gives the branch of solutions of  
(\ref{IM2}) through $(0, t_0)$, and $t(1)$ will be 
the solution to the original problem (\ref{interIfixplusC}) 
(method of imbedding or continuity \cite{Mey}).

By the implicit function theorem, if grad $\Psi$ is continuous and 
$(\partial \Psi/\partial t)(0,t_0) \not=0$, there exists a branch 
$t(\lambda)$ through $(0,t_0$) on a sufficiently small interval around 
$\lambda=0$. One then extends $t(\lambda)$ by a standard argument.  
Denoting partial derivatives by 
subscripts following a comma, one has 
\beq
t_{,\,\lambda} = -{\Psi_{,\,\lambda} (\lambda,t) \over 
\Psi_{,\, t} (\lambda,t)}  \; .\label{IM3}
\eeq
For sufficiently small $\Delta \lambda $ then, and by the mean value 
theorem, one can write  
\[ t(0+\Delta \lambda) = t(0) + t_{,\, \lambda}(\xi \Delta \lambda) \,\Delta 
\lambda = t(0) + t_{,\, \lambda}(0) \,\Delta \lambda 
+ O(\Delta\lambda^2) \]
for some $0 <\xi <1$. One can then use (\ref{IM3}) to find 
$t_{,\,\lambda}(0+\Delta \lambda)$, and repeat the procedure to 
obtain $t(0+2\Delta \lambda)$, and so on. If 
grad $\Psi$ is well-behaved throughout the relevant $\lambda - t$ domain,  
this procedure constructs the desired branch away from 
the initial point as long as $\Psi_{,\, t}\not=0$ along the branch. 
The existence of a solution $t(1)$ is therefore 
guaranteed by basic existence theorems (see e.g. \cite{Mey}) 
if this condition is satisfied throughout the interval 
$0\leq \lambda \leq 1$. 
 
Now  
\bea 
\Psi_{,\,t}(\lambda, t) & = & 
\left[ - {h(\alpha_h, t)_{,\,\alpha} + 
\lambda\, A^+(\alpha_h)
\over h(\alpha_h,t)_{,\,\alpha}
+ A(\alpha_h) }  +  1 \right]
h(\alpha_h, t)_{,\,t}  \nonumber \\
& < & 0 \quad \hspace{7cm} 0\leq \lambda \leq 1  \label{IM4}
\eea
by (\ref{A>A+}) and since $h(\alpha_h, t)_{,\,t} <0$. 
On the other hand, 
\beq
\Psi_{,\,\lambda}(\lambda, t) = \Big[\, \dPhi{n}^+(\alpha_h(t)) - \dPhi{n}^+(
\alpha_h^+(t_I))\,\Big] \; >  \; 0 \;, 
\qquad 0\leq \lambda \leq 1\; .\label{IM5}
\eeq 
by (\ref{alphtzero1}) and IV.5. 
Thus, from (\ref{IM3}), $t(\lambda)$ is an increasing function of 
$\lambda$, and extends to the solution $t(1) > t_0$.
 
Conversely, if there is a $t_I$ such that 
$\alpha_h(t_I) < \alpha_h^+(t_I)$, one can find  $t_0$ 
such that (\ref{htzero}) is satisfied (Figure \ref{C1abfig}(b)), 
where now $t_0 < t_I$ and 
\beq
\alpha_h^+(t_I) > \alpha_h^+(t_0) > \alpha_h(t) \;, \qquad 
t\leq t_0 \,. \label{alphtzero2}
\eeq
Now (\ref{IM4}) remains unchanged, but 
\beq 
\Psi_{,\,\lambda}(\lambda, t) = \Big[\, \dPhi{n}^+(\alpha_h(t)) - \dPhi{n}^+(
\alpha_h^+(t_I))\,\Big] \; <   \; 0 \;, 
\qquad 0\leq \lambda \leq 1\; ,\label{IM6}
\eeq 
by (\ref{alphtzero2}) and IV.5. 
It follows that 
$t(\lambda)$ is now a decreasing function of $\lambda$, 
i.e. if $\alpha_h(t_I) < \alpha_h^+(t_I)$ at the starting point 
one moves backwards in $t$ to hit the point $t(1)$ where 
$\alpha_h(t(1))=\alpha_h^+(t(1))$ and (\ref{interIfixplusC}) is 
satisfied. 
This concludes the demonstration of the existence of a solution to 
(\ref{interIfixplus}).

\noindent{\bf \S2.} Going back to (\ref{ratio3}),  
assume there is a $t_I$ such that $\alpha_{\Lambda,\,h}^{(n)}(t_I) < 
\alpha_{\Lambda,\,h}^{+(n)}(t_I)$. Then,  
setting $t=t^+=t_I$, one has 
\bea
\left(\,1+ {Z_\Lambda^{(-)} \over Z_\Lambda }\,\right) & = & 
{ 2 \tilde{Z}^+_{\Lambda^{(n)}}\,(\beta, h,  
\alpha_{\Lambda,\, h}^{+(n)}(t_I),\, t_I) \over 
\tilde{Z}_{\Lambda^{(n)}}\,(\beta, h,  
\alpha_{\Lambda,\,h}^{(n)}(t_I),\, t_I) } \nonumber \\
 & \geq & 
\left(\,1+ 
{ Z_{\Lambda^{(n)}}^{(-)}\,\Big(\{\,\tilde{c}_j(n,\alpha_{\Lambda,\,h}^{(n)}
(t_I))\,\}\Big)
\over Z_{\Lambda^{(n)}}\,\Big(\{\,\tilde{c}_j(n,\alpha_{\Lambda,\,h}^{(n)}
(t_I))\,\}\Big) } \,\right) \\ \label{ratiolowerC1}
& \geq & 
\left(\,1+ { Z_{\Lambda^{(n)}}^{(-)}\,\Big(\{\,c^U_j(n)\,\}\Big)  
\over Z_{\Lambda^{(n)}}\,\Big(\{\,c^U_j(n)\,\}\Big) } \,\right) 
\label{ratiolowerC2}
\eea
by IV.5 and (\ref{ratioder}). On the other hand an upper bound is 
always obtained by using (\ref{Z>Z-}).  
Taking the coefficients 
$\tilde{c}_j(n,\alpha)$ in the form (\ref{interc1}) with (\ref{lowerc3}), 
which interpolate between this upper bound at $\alpha=0$ and 
the lower bound (\ref{ratiolowerC2}) at $\alpha=1$, one  obtains 
\beq 
1={ Z_{\Lambda^{(n)}}^{(-)}\,\Big(\{\,c^L_j(n)\,\}\Big)  
\over Z_{\Lambda^{(n)}}\,\Big(\{\,c^L_j(n)\,\}\Big) } 
\geq {Z_\Lambda^{(-)} \over Z_\Lambda } \geq 
{ Z_{\Lambda^{(n)}}^{(-)}\,\Big(\{\,c^U_j(n)\,\}\Big)  
\over Z_{\Lambda^{(n)}}\,\Big(\{\,c^U_j(n)\,\}\Big) } 
\label{ratiolowerupperC} \, .
\eeq 
This is (\ref{ratiolowerupper}) again (for the case $c^L_j(n)=0$).  
It follows from (\ref{ratiolowerupperC}) that there exist 
a value $\alpha_\Lambda^{*\,(n)}$ such that (\ref{ratio6}) holds. 
This value is unique by monotonicity from (\ref{ratioder}).

This is an alternative way of treating the $\alpha_h(t_I) < \alpha_h^+(t_I)$ 
case in \S1 above. 

(\ref{ratiolowerC1}) - (\ref{ratiolowerupperC}) partially 
implement the alternative approach to (\ref{ratio6}) outlined in 
the last paragraph of section \ref{Z-/Z}. One way to complete it would be 
to show that, for some interpolation $h$, there is
at least  one value $t_I$ such that  
$\alpha_{\Lambda,\,h}^{(n)}(t_I) < \alpha_{\Lambda,\,h}^{+(n)}(t_I)$.

\section{Appendix} 
\setcounter{equation}{0}

From (\ref{RG5}) with integer $\zeta >  1$ one has 
\beq
\hat{F}_i(n+1) = \delta_{0,i} + (1-\delta_{0,i})\,\zeta \,c_i(n)  + 
\sum_{k=2}^\zeta\; {\zeta\choose k} 
\,I_i(k) \;, \label{Fhat1}
\eeq
where 
\beq
I_i(k) = {1\over d_i}\sum_{\{j_s| \;0 < s\leq k\}}
d_{j_1}c_{j_1}(n)  
\cdots d_{j_k}c_{j_k}(n)\sum_{l_1, \cdots,\, l_{k-1}}
\,
\Delta(j_1,j_2,l_2) \Delta(l_2,j_3,l_3) \cdots \Delta(l_{k-1}, j_k,i) 
\label{Idef}
\eeq 
with $\Delta(j,k,l)=1$ if $j,k,l$ form the `angular momentum addition 
triangle' relation, i.e. $l= |j-k|, \cdots, j+k$, and $0$ otherwise.  
Then 
\bea 
\sum_{i\not=0} d_i I_i(k)  & = & \sum_{\{j_s| \;0 < s\leq k\}}
d_{j_1}c_{j_1}(n) \cdots d_{j_k}c_{j_k}(n)\sum_{i\not=0,\, l_1, \cdots,\, 
l_{k-1}}
\,
\Delta(j_1,j_2,l_2) \Delta(l_2,j_3,l_3) \cdots \Delta(l_{k-1}, j_k,i) 
\nonumber \\
& \leq & \sum_{\{j_s| \;0 < s\leq k\}} d_{j_1}c_{j_1}(n)  
\cdots d_{j_k}c_{j_k}(n) \,d_{j_2} d_{j_3} \cdots d_{j_k} 
\nonumber \\
& \leq & ||g(n)||^k\;.  \label{Isum}
\eea 
Hence, with $b\geq 2$, 
\bea 
\sum_{i\not=0} d_i^2\hat{F}_i(n+1)^{b^2} & = & 
\sum_{i\not=0} d_i^2 \left[\,\zeta \,c_i(n) + \sum_{k=2}^\zeta\; 
{\zeta\choose k} \,I_i(k) \,\right]^{b^2} \nonumber \\
& \leq &  
\left[\, \zeta \sum_{i\not=0} d_i^2 c_i(n) + 
\sum_{k=2}^\zeta\; {\zeta\choose k} 
\,\sum_{i\not=0} d_i I_i(k) \,\right]^{b^2}\nonumber \\
& \leq & 
\left[ \,\zeta ||g(n)|| + \sum_{k=2}^\zeta\; 
{\zeta\choose k} \,||g(n)||^k\,\right]^{b^2}
\nonumber \\
& = & 
\left[\,\left[\,1+ ||g(n)||\,\right]^{\,\zeta} -1\,\right]^{b^2} 
\nonumber \\
& \leq & \Big[\,\zeta\; ||g(n)||\,\Big]^{b^2} 
\,\Big[\,1+ ||g(n)||\,\Big]^{(\zeta-1) \,b^2} \;. \label{Fsum}
\eea
Also, from (\ref{Fhat1}), (\ref{Idef}) 
\bea
\hat{F}_0(n+1) \geq 1+ {\zeta (\zeta-1)\over 2} I_0(2) 
& = & 1 +{\zeta (\zeta-1)\over 2} \sum_{j\not= 0} d_j^2 c_j(n)^2 
\label{F0lower1D}\\
& > & 1 \;, \label{F0lower2D} 
\eea
whereas also from (\ref{RG5}) 
\beq
\hat{F}_0(n+1)\leq  \Big[\,1 + ||g(n)||\,\Big]^\zeta \;.\label{F0upper}
\eeq
Combining (\ref{Fsum}) and (\ref{F0lower2D}) and taking $r=1$ gives 
(\ref{gnormrecur}). 

For $r\not=1$, (\ref{Fsum}) and, hence,  
(\ref{gnormrecur}) hold with the replacement $b^2 \to b^2r$,  
provided $b^2 r > 2$.

\end{document}